\documentclass[12pt]{article}
\usepackage[margin=1in]{geometry} 
\usepackage{amsmath,amsthm,amssymb,amsfonts,mathtools,bm,mathdots}
\usepackage[makeroom]{cancel}
\usepackage{graphicx}
\usepackage{comment}
\usepackage{fancyhdr}
\usepackage[bottom]{footmisc}
\usepackage{centernot}
\usepackage{tocdata}
\usepackage{lipsum}
\usepackage{multirow}
\usepackage{float}
\usepackage{bbold}
\usepackage{listings}
\usepackage{color} 
\definecolor{mygreen}{RGB}{28,172,0} 
\definecolor{mylilas}{RGB}{170,55,241}
\usepackage{array}
\usepackage{tabto}
\usepackage{wrapfig}
\usepackage{caption}
\usepackage{subcaption}
\newcolumntype{P}[1]{>{\centering\arraybackslash}p{#1}}
\newcolumntype{M}[1]{>{\centering\arraybackslash}m{#1}}
\usepackage[colorlinks=true, allcolors=blue]{hyperref}
\usepackage[shortlabels]{enumitem}

\usepackage{amsfonts}
\usepackage{indentfirst}
\usepackage{mathrsfs}
\usepackage{hyperref}
\usepackage{xcolor}
\usepackage{amssymb}
\usepackage{tikz}
\usepackage{tkz-graph}
\usepackage{bm}
\usepackage{multirow}
\usepackage{color}
\usetikzlibrary{decorations.pathreplacing}
\usepackage{diagbox}
\usepackage[toc,page]{appendix}

\usepackage{cite}

\usepackage[autostyle=true]{csquotes} 

\newtheorem{theorem}{Theorem}[section]
\newtheorem*{theorem*}{Theorem}

\newtheorem{proposition}[theorem]{Proposition}
 
\newtheorem{conjecture}[theorem]{Conjecture} 

\theoremstyle{definition}
\newtheorem{definition}[theorem]{Definition}
\newtheorem{example}[theorem]{Example}

\newtheorem{problem}[theorem]{Problem}

\newtheorem{remark}[theorem]{Remark}

\DeclareMathAlphabet{\mathpzc}{OT1}{pzc}{m}{it}

\DeclareMathOperator{\Ric}{Ric}

\renewcommand{\headrulewidth}{0.4pt}
\renewcommand{\footrulewidth}{0.4pt}

\begin{document}
\title{Lectures on the Calabi-Yau Landscape}
\author{Jiakang Bao$^{1,2}$\footnote{$\mathtt{jiakang.bao18@imperial.ac.uk}$}, Yang-Hui He$^{1,3,4}$\footnote{$\mathtt{hey@maths.ox.ac.uk}$}, Edward Hirst$^{1,2}$\footnote{$\mathtt{Edward.Hirst@city.ac.uk}$}, Stephen Pietromonaco$^{5}$\footnote{$\mathtt{spietro@math.ubc.ca}$}}
\date{\footnotesize{\textit{$^1$Department of Mathematics, City, University of London, EC1V 0HB, UK\\
$^2$Department of Physics, Imperial College London, SW7 2AZ, UK\\
$^3$Merton College, University of Oxford, OX14JD, UK\\
$^4$School of Physics, NanKai University, Tianjin, 300071, P.R. China\\
$^5$Department of Mathematics, University of British Columbia, V6T 1Z2, Canada}}}
\maketitle

\begin{abstract}
In these lecture notes, we survey the landscape of Calabi-Yau threefolds, and the use of machine learning to explore it. We begin with the compact portion of the landscape, focusing in particular on complete intersection Calabi-Yau varieties (CICYs) and elliptic fibrations. Then we examine non-compact Calabi-Yau manifolds which are manifest in Type \uppercase\expandafter{\romannumeral2} superstring theories. They arise as representation varieties of quivers, used to describe gauge theories in the bulk familiar four dimensions. Finally, given the huge amount of Calabi-Yau data, whether and how machine learning can be applied to algebraic geometry and string landscape is also discussed. These notes are directed to the beginning graduate student interested in mathematics and in physics, and are based on lectures given by the $2^\text{nd}$ author at the 2019 PIMS Summer School on Algebraic Geometry in High-Energy Physics at the University of Saskatchewan.
\end{abstract}

\renewcommand{\headrulewidth}{0.4pt}
\renewcommand{\footrulewidth}{0.4pt}

\setcounter{page}{1}
\numberwithin{equation}{section}

\newpage
\tableofcontents
\clearpage

\section{Introduction}

Superstring theories demand our spacetime dimension to be 10, which means we should reduce them to an effectively 4-dimensional theory. The standard solution of string compactification, as a generalization of Kaluza-Klein compactification, renders the extra six dimensions Calabi-Yau (CY). Thus, the study of Calabi-Yau and algebraic geometry has entered the field of theoretical physics.

In order to avoid an excess of symmetries in our observed 4-dimensional universe, isometries in our geometry, which leads to extra graviphotons, is not allowed \cite{Douglas:2015aga}. This leaves us the only option of manifolds of complex dimension 3, which requires K\"ahler structure and vanishing first Chern classes ($c_1=0$). As will be explained in \S\ref{subsec:CYcompST}, we also want the manifold to be Ricci-flat. However, given a K\"ahler manifold with zero $c_1$, the existence of a (unique) K\"ahler metric in the same K\"ahler class with vanishing Ricci form is not self-evident. Followed by the work of Calabi \cite{Calabi} and Yau \cite{Yau1, Yau2}, mathematicians reached a great success in studying CY manifolds. Later, physicists realized the crucial role CY manifolds play in fundamental physics as aforementioned. Discoveries in physics enabled people to reconstruct the Standard Model from compactifications and also led to the mirror symmetry which is now a focused interface of mathematics and physics \cite{CHSW}. More details and discussions on the physcial predictions from CY manifolds can be found in \cite{Douglas:2015aga}. Nowadays, thanks to the information age, we are able to let machines help us learn the structure of CY manifolds due to the large volume of data which has been compiled since the mid-1980s by physicists and mathematicians. This even brings computer science and data science into this interdisciplinary area.

The outline is organized as follows. In Part \ref{part1}, we mainly focus on compact CY landscape. We start with a background on Calabi-Yau geometry. We also pay our attention to the complete intersection Calabi-Yaus (CICYs). Then we contemplate the non-compact case in Part \ref{part:NCCYL}. In this part, more physics and mathematics, such as quivers and toric varieties, and their relations are discussed. Finally, we apply machine learning to the study of CY landscape in Part \ref{machinelearning}. Along with a quick introduction to machine learning, we perform this technique to different topics in mathematics. In the appendices, some prerequisites are provided.

\clearpage
\part{Compact Calabi-Yau Landscape}\label{part1}

Some basic topological or geometric facts are given in Appendix \ref{complexgeometry}. For far more detailed treatment on what follows, we refer the reader to \cite{hubsch_calabi-yau_1992,CYLandscape, deeplearninglandscape,PLBpaper}.

\section{Calabi-Yau Geometry in Math and Physics}\label{sec:CYGinMP}

The story of Calabi-Yau manifolds originates in the mid-1950s with the following conjecture of Eugenio Calabi.  

\begin{conjecture}{\bfseries(The Calabi Conjecture)}
Let $(X,g, \omega)$ be a compact K\"{a}hler manifold, and fix $R \in \Omega^{1,1}(X)$ such that $[R] = c_{1}(T_{X}) \in H^{1,1}(X)$.  Then there exists a unique K\"{a}hler metric $\widetilde{g}$ with K\"{a}hler form $\widetilde{\omega}$ such that $[\omega] = [\widetilde{\omega}]$, and 
\begin{equation*}
R = \Ric(\widetilde{\omega})
\end{equation*}
where $\Ric(\widetilde{\omega})$ is the Ricci form of $\widetilde{\omega}$.\label{calabiconjecture}  
\end{conjecture}

\noindent The power of this conjecture is that it describes complicated geometric data (curvature) in terms of simpler topological data (Chern classes).  For example, in complex dimension 1, this conjecture reduces to the Gauss-Bonnet theorem for Riemann surfaces, which says that the curvature is determined completely by the genus.  In higher dimensions, the conjecture is that the curvature is controlled by the first Chern class (of the tangent bundle).  

Calabi himself proved the uniqueness part of his conjecture, but the existence remained an open problem for 20 years before Shing-Tung Yau completed the proof, for which he received the Fields Medal in 1982.  

\begin{theorem}{\bfseries(Yau)}
The Calabi conjecture holds.   
\end{theorem}

We will be primarily interested in the special case of $R=0$, in which we say that $X$ admits a \emph{Ricci-flat} metric.  In general relativity, Riemannian manifolds with Ricci-flat metrics are vacuum solutions of Einstein's equations (that is, solutions without matter and energy).  We are therefore interested in such manifolds which are K\"{a}hler.  This leads us to the definition of a Calabi-Yau manifold\footnote{In fact, the word ``Calabi-Yau'' was coined by physicists later \cite{CHSW} for Ricci-flat K\"ahler manifolds.}.

\begin{definition}
Let $X$ be a compact K\"{a}hler manifold with $\dim_{\mathbb{C}}(X) = n$.  We say $X$ is a Calabi-Yau $n$-fold if it admits a Ricci-flat metric\footnote{Yau's proof of the Calabi conjecture was not constructive, and to-date, there is not a single compact Calabi-Yau manifold where the Ricci-flat metric is known explicitly (outside of trivial cases of tori).  This is an important open problem.} of strictly SU($n$) holonomy.  
\end{definition}

\noindent Let us give some low-dimensional examples of Calabi-Yau manifolds: 
\begin{enumerate}
\item The only Calabi-Yau manifold of (complex) dimension 1 is an elliptic curve.  Thus, there is a single topological type.  

\item The Calabi-Yau manifolds of complex dimension 2 are called K3 surfaces.  A simple construction is as a smooth quartic hypersurface in $\mathbb{P}^{3}$.  All K3 surfaces are simply connected, and diffeomorphic to one another; so there is only one topological type.  (Note that 4-dimensional tori are indeed Ricci flat, but they do not satisfy the condition on the holonomy group in the definition.)  
\end{enumerate}

\begin{proposition}\label{proppy:CYeqcond}
For $X$ as in the definition, the following are equivalent\footnote{There are some subtleties in these propositions. The second one is actually weaker. For instance, complex tori with dimension greater than one have vanishing first Chern classes, but they fail to satisfy the fifth one. On the other hand, people often count these as Calabi-Yaus as they have trivial holonomies and infinite fundamental groups. Moreover, we also have non-algebraic K3 surfaces that fail the fifth condition even though they are simply connected with holonomy SU(2) \cite{math/0108088}. Anyway, people adopt different definitions in different literature. This won't be an issue in our applications.}:  

\begin{enumerate}
\item $X$ is a Calabi-Yau $n$-fold.  

\item The first Chern class of $X$ vanishes; $c_{1}(T_{X})=0$.  

\item There exists a covariantly constant spinor on $X$.  

\item There exists a non-vanishing holomorphic $n$-form on $X$.  

\item $X$ is a smooth projective algebraic variety with trivial canonical line bundle $\omega_{X} \cong \mathcal{O}_{X}$, where $\omega_{X} = \bigwedge^{n} T_{X}^{*}$, and which additionally satisfies $H^{k}(X, \mathcal{O}_{X})=0$ for $0< k < n$.  
\end{enumerate}
\end{proposition}

The final characterization in the proposition is clearly the preferable one in algebraic geometry.  We can remove the hypothesis of projectivity, which results in non-compact Calabi-Yau manifolds, of interest to us in Part \ref{part:NCCYL}.  We could also allow for mild singularities, which inevitably arise when studying families of Calabi-Yau manifolds.  

\begin{remark}\label{imprmkk}
One must beware of mildly different definitions of Calabi-Yau.  Our definition excludes all tori (in particular, abelian varieties) and, for example, the threefold K3$\times E$; the product of a K3 surface and an elliptic curve.  These spaces admit Ricci-flat metrics, though of holonomy strictly contained in SU($n$).  In physics, this will translate into the low-energy theory having enhanced supersymmetry.  Both abelian threefolds and K3$\times E$ are of interest in enumerative geometry.
\end{remark}

\subsection{Topological Data}  

One can assign to a complex manifold $X$ the Hodge cohomology groups
\begin{equation*}
H^{p,q}(X) \coloneqq H^{q}(X, \Omega_{X}^{p})
\end{equation*}
with Hodge numbers $h^{p,q}$ the corresponding dimensions.  If $X$ is compact and K\"{a}hler, the topological Euler characteristic is given by
\begin{equation}\label{eqn:HodgeEulChar}
\chi(X) = \sum_{p,q=1}^{\dim X} (-1)^{p+q}h^{p,q}.
\end{equation}
If $X$ is a compact Calabi-Yau threefold, then due to various symmetries \cite{CYLandscape,deeplearninglandscape,PLBpaper} the only relevant Hodge numbers are $h^{1,1}$ and $h^{2,1}$.  By Proposition \ref{proppy:CYeqcond}, $X$ is a smooth projective variety with vanishing $h^{1,0}, h^{2,0}$ and therefore by the Hodge decomposition 
\begin{equation*}
H^{1,1}(X) \cong H^{2}(X, \mathbb{C}).
\end{equation*}
We can choose an integral basis $\{J_{k}\}_{k=1, \ldots, h^{1,1}}$ of $H^{2}(X, \mathbb{C})$ such that the K\"{a}hler cone is $\mathcal{K} = \big\{ \sum_{k} t_{k}J_{k} \, \big| \, t_{k} \in \mathbb{R}_{>0} \big\}$.  In other words, the quantity $h^{1,1}$ measures the number of K\"{a}hler classes on $X$ (or by dualizing, the number of curve classes).  Using the Calabi-Yau condition, we similarly have
\begin{equation*}
H^{2,1}(X) \cong H^{1}(X, T_{X}).
\end{equation*}
The cohomology group on the right encodes the infinitesimal deformations of the complex/algebraic structure of $X$.  

Therefore on a Calabi-Yau threefold, the Hodge number $h^{2,1}$ measures the dimension of the space of complex/algebraic deformations, while $h^{1,1}$ measures the dimension of the K\"{a}hler cone.  The two Hodge numbers determine the topological Euler characteristic via (\ref{eqn:HodgeEulChar})
\begin{equation}
\chi(X) = 2(h^{1,1} - h^{2,1}).
\end{equation}

Using the chosen basis of $\mathcal{K}$ we define the triple intersection form of $X$
\begin{equation*}
d_{rst} = \int_{X} J_{r} \wedge J_{s} \wedge J_{t}.  
\end{equation*}
This integral can be hard to compute in general, but we can use the following result \cite[Thm. 1.3]{hubsch_calabi-yau_1992}.  If we have an embedding $f: X \hookrightarrow A$ with $A$ a smooth projective variety of dimension $m+3$, then for all $\omega \in H^{k}(A)$
\begin{equation*}
\int_{X} \omega|_{X} = \int_{A} \omega \wedge \eta
\end{equation*}
where $\eta$ is a $(m,m)$-form which when restricted to $X$ is the top Chern class of the normal bundle $\mathcal{N}_{X/A}$.  For our purposes, $A$ will be a simpler space than $X$ itself; for example, a projective space or product of projective spaces.  

For any K\"{a}hler threefold, the total Chern class can be written in the chosen basis of $\mathcal{K}$ as
\begin{equation*}
c(T_{X}) = 1 + \sum_{r=1}^{h^{1,1}}[c_{1}(T_{X})]_{r}J_{r} + \sum_{r,s=1}^{h^{1,1}}[c_{2}(T_{X})]_{rs}J_{r} \wedge J_{s} + \sum_{r,s,t=1}^{h^{1,1}}[c_{3}(T_{X})]_{rst}J_{r} \wedge J_{s} \wedge J_{t}.
\end{equation*}
Moreover, the topological Euler characteristic of a K\"{a}hler manifold $X$ is the integral over $X$ of the top Chern class of $T_{X}$.  Using the triple intersection form, we can therefore express
\begin{equation*}
\chi(X) = \sum_{r, s, t =1}^{h^{1,1}} d_{rst} [c_{3}(T_{X})]_{rst}.
\end{equation*}
For a Calabi-Yau threefold, of course $c_{1}(T_{X})=0$, so that leaves $c_{2}(T_{X})$ to be independently specified.  

\begin{theorem}[\bfseries Wall]\label{wall}
The topological type of a compact Calabi-Yau threefold is completely determined by the Hodge numbers $h^{p,q}$, the triple intersection form $d_{rst}$, and the second Chern class $c_{2}(T_{X})$.  
\end{theorem}
It is convenient to contract $c_{2}$ with $d$ by defining $[c_{2}(T_{X})]_{r} \coloneqq \sum_{s,t}[c_{2}(T_{X})]_{rs}d_{rst}$.  It suffices to record this contraction instead of the individual components $[c_{2}(T_{X})]_{rs}$.  Therefore, by Theorem \ref{wall}, the data determining the topological type of a Calabi-Yau threefold is:
\begin{equation}\label{eqn:CY3Data}
\bigg\{ (h^{1,1},h^{2,1}), \,\,\, [c_{2}(T_{X})]_{r}, \,\,\, d_{rst}    \bigg\} \,\,\,\,\,\,\,\,\,\, r,s,t = 1, \ldots, h^{1,1}.
\end{equation}

Recall from Section \ref{sec:CYGinMP} that for Calabi-Yau manifolds of dimensions 1 and 2, there is respectively a single topological type.  Does this pattern persist in dimension 3?  Spectacularly, no.  The lower bound on the number of topological types of Calabi-Yau threefolds is currently around 500,000,000!  But there is the following conjecture.
\begin{conjecture}[\bfseries Yau]
The number of topological types of Calabi-Yau threefolds is finite\footnote{In fact, this conjecture is made for any CY $n$-folds. It is certainly true for $n=1,2$.}.
\end{conjecture} 
\noindent In other words, there are finite possibilities for the values in the data set (\ref{eqn:CY3Data}).

\begin{remark}
Beware that even after fixing the topological type of the Calabi-Yau, there is still generally a moduli of algebraic/complex structures on the variety of fixed type.  This is typical of moduli problems: specify as much discrete data as possible, which fixes the topological type, and then study families of complex structures.  
\end{remark}

\subsection{String Compactifications}\label{subsec:CYcompST}

Calabi-Yau threefolds entered physics through string theory in the late 80s.  The consistency of the physical string theories (Type I, Type IIA, Type IIB, and the Heterotic theories) remarkably requires that the (real) dimension of spacetime be 10.  So we obviously have to contend with the fact that we only observe 4 dimensions.  The idea behind \emph{string compactifications} is to decompose the 10-dimensional spacetime $M_{10}$ as
\begin{equation}\label{strngcompp}
M_{10} = M_{4} \times X
\end{equation}
where $M_{4}$ is our 4-dimensional spacetime, and $X$ is a compact 6-dimensional manifold.  The vague intuition should be that the extra 6 dimensions of $X$ are tightly curled-up and unobservable at small energies.  

If $X$ is a complex threefold, then it has real dimension 6.  But why do we want $X$ to additionally be Calabi-Yau?  It is because Calabi-Yau manifolds are those admitting Ricci-flat metrics.  In general relativity Ricci-flat manifolds correspond to a vacuum configuration of spacetime, i.e. a universe without matter or energy.  Therefore, compactifying on a Calabi-Yau threefold $X$, as in (\ref{strngcompp}), models a string theory vacuum.   

Let us tie this back in with our exploration of the Calabi-Yau landscape.  The vague principle one should keep in mind is:
\begin{center}
\fbox{\begin{minipage}{28em}
As $X$ varies over the compact Calabi-Yau landscape, the physics observed in $M_{4}$ changes.  In other words, the topology and geometry of $X$ dictates physical phenomena in spacetime.
\end{minipage}}
\end{center}

\section[C.I.C.Y.]{Complete Intersections in Products of Projective Spaces (CICYs)}

In this section we begin constructing our first examples of compact Calabi-Yau threefolds.  The simplest (and most famous) Calabi-Yau threefold is the quintic, and more generally, the cyclic manifolds.  Subsuming these examples, is the important class of complete intersection in products of projective spaces, or CICY for short.  After constructing these geometries, we show how certain crucial topological information is encoded into the defining equations.    

\subsection{Cyclic Calabi-Yau Threefolds}

Let us now construct the most straightforward example of a Calabi-Yau manifold in each dimension.  Let $f(x_{0}, \ldots, x_{n})$ be a homogeneous degree $d$ polynomial, or equivalently, a section of the line bundle $\mathcal{O}_{\mathbb{P}^{n}}(d)$.  The vanishing locus of the section defines a degree $d$ hypersurface $X$ in the projective space $\mathbb{P}^{n}$.

\begin{theorem}{\bfseries (The Adjunction Formula)}
Let $X \subset \mathbb{P}^{n}$ be a smooth, closed subvariety of codimension $m$.  The canonical bundle of $X$ is given by
\begin{equation}
\omega_{X} = \Lambda^{m} \mathcal{N}_{X/\mathbb{P}^{n}} \otimes_{\mathcal{O}_{X}} \mathcal{O}_{\mathbb{P}^{n}}(-n-1)\big|_{X}
\end{equation}
where $\mathcal{N}_{X/\mathbb{P}^{n}}$ is the normal bundle of $X$ in $\mathbb{P}^{n}$ \cite{hartshorne}.  
\end{theorem}

Since $X$ is a divisor cut out by a section of $\mathcal{O}_{\mathbb{P}^{n}}(d)$, the normal bundle is the line bundle $\mathcal{N}_{X/\mathbb{P}^{n}} = \mathcal{O}_{\mathbb{P}^{n}}(d)|_{X}$.  Therefore, the canonical bundle will be trivial if and only if $d=n+1$.  By the Lefschetz hyperplane theorem, $\pi_{1}(X)$ is trivial.  We have therefore shown the following.  

\begin{proposition}
A homogeneous polynomial of degree $n+1$ in the $n+1$ projective coordinates on $\mathbb{P}^{n}$ defines a compact Calabi-Yau $n$-fold as a divisor $X \subset \mathbb{P}^{n}$.  
\end{proposition}

Since we are interested in dimension 3, of most importance here will be the \emph{the quintic} Calabi-Yau threefold constructed from a quintic polynomial in $\mathbb{P}^{4}$.  For example, the Fermat quintic is the vanishing locus of 
\begin{equation}\label{eqn:FermatQuintic}
f(x_{0}, x_{1}, x_{2}, x_{3}, x_{4}) = x_{0}^{5} + x_{1}^{5} + x_{2}^{5} + x_{3}^{5} + x_{4}^{5}.
\end{equation}

\begin{remark}
Note that saying ``the" quintic is somewhat misleading, as we actually get a family of Calabi-Yau threefolds, by varying the coefficients in the quintic polynomial.  However, these correspond to various complex structures on the same underlying topological type.  It is conventional to refer to the entire family as "the quintic."  Similarly, note that certain quintic polynomials give singular varieties.  Unless mentioned otherwise, we will assume to be working with a smooth member of the family, for example (\ref{eqn:FermatQuintic}).  
\end{remark}

How can we generalize the quintic Calabi-Yau?  The quintic is a hypersurface, and the most immediate generalization of a hypersurface is a complete intersection $X \subset \mathbb{P}^{n}$, which means the codimension of $X$ equals the number of polynomials cutting it out.  This is the most ideal intersection, though is quite rare in the world of varieties.  

Suppose we have $k$ homogeneous polynomials $\{f_{i}\}_{i=1, \ldots, k}$ on $\mathbb{P}^{n}$ with $q_{i} \in \mathbb{Z}_{\geq 0}$ the degree of $f_{i}$.  The vanishing locus of the $f_{i}$ produces a compact Calabi-Yau threefold as a complete intersection in $\mathbb{P}^{n}$ if
\begin{equation}\label{eqn:cyclicCYconds}
\begin{split}
& k=n-3 \,\,\,\,\,\,\,\,\,\,\,\,\,\,\,\,\,\,\,\,\,    (\text{Complete intersection condition}) \\
& n+1 = \sum_{i=1}^{k} q_{i} \,\,\,\,\,\,\,\,\,\,\,\,\,\, (\text{Generalization of Adjuntion})
\end{split}
\end{equation}
One can show the fundamental group is trivial using a generalization of the Lefschetz hyperplane theorem \cite[Thm. 1.4]{hubsch_calabi-yau_1992}.  We call such a manifold a \emph{cyclic} Calabi-Yau threefold.  A notation which will prove helpful in the following section is to denote a collection of degrees as
\begin{equation*}
M= [
\begin{array}{cccc}
q_{1} & q_{2} & \cdots & q_{k} \\
\end{array}
]
\end{equation*}
with $X_{M}$ the corresponding cyclic Calabi-Yau.  Note that $n$ can be recovered from the condition $n= k +3$.  

Clearly, (\ref{eqn:cyclicCYconds}) defines a rather constrained combinatorial problem, and it turns out there are only 5 solutions.  In the notation above, these are:
\[ [\, 5 \,], \,\,\,\,\,\,\,\, [ \, 2 \,\,\, 4 \,], \,\,\,\,\,\,\,\,  [ \, 3 \,\,\, 3 \,], \,\,\,\,\,\,\,\, [\, 3 \,\,\, 2 \,\,\, 2 \,], \,\,\,\,\,\,\,\, [ \, 2 \,\,\,2 \,\,\,2 \,\,\, 2 \,].\]
The first example is the quintic, the second example is the complete intersection of a quadric and a quartic in $\mathbb{P}^{5}$, the third example is the complete intersection of two cubics in $\mathbb{P}^{5}$, etc.

\subsection{CICY Calabi-Yau Threefolds}\label{CICY3} 

We can achieve a far greater generalization of the quintic by considering complete intersections in not just the ambient space $\mathbb{P}^{n}$, but rather in a product of projective spaces
\begin{equation*}
A  = \mathbb{P}^{n_{1}} \times \cdots \times \mathbb{P}^{n_{m}}. 
\end{equation*} 
Suppose we have $k$ multi-homogeneous polynomials $\{f_{i}\}_{i=1, \ldots, k}$ on $A$, with multi-degrees $q^{i}_{j} \in \mathbb{Z}_{\geq 0}$ where $i=1, \ldots, k$ and $j=1, \ldots, m$.  In words, $q^{i}_{j}$ is the degree of the $i$-th polynomial on the $j$-th factor of $A$.  Generalizing the notation for cyclic Calabi-Yau threefolds, we package the data into the configuration matrix
\begin{equation}\label{eqn:confmatr}
M=\left[
\begin{array}{cccc}
q^{1}_{1} & q^{2}_{1} & \cdots & q^{k}_{1} \\
q^{1}_{2} & q^{2}_{2} & \cdots & q^{k}_{2} \\
\vdots & \vdots & \ddots & \vdots \\
q^{1}_{m} & q^{2}_{m} & \cdots & q^{k}_{m}\\
\end{array}
\right]
\end{equation}
We define $X_{M} \subset A$ to be the vanishing locus of the $\{f_{i}\}_{i=1, \ldots, k}$.  The projective variety $X_{M}$ is a Calabi-Yau threefold if the following conditions hold
\begin{equation}\label{eqn:CICYconds}
\begin{split}
& \,\,\,\,\,\,\,\,\,\,\,\,\,\,\,\,\,\,\,\,\,\,\,\,\,\,\,\,\,\,\,\,\,\,\,\,\,\,   k= \sum_{i=1}^{m} n_{i} -3 \,\,\,\,\,\,\,\,\,\,\,\,\,\,\,\,\,\,\,\,\,\,\,\,\,    (\text{Complete intersection condition}) \\
& n_{j}+1 = \sum_{i=1}^{k} q^{i}_{j}, \,\,\, \text{for all} \,\,\, j=1,\ldots,m \,\,\,\,\,\,\,\,\,\,\,\,\,\, (\text{Generalization of Adjuntion})
\end{split}
\end{equation}
Such a $X_{M}$ is called a \emph{CICY}, which refers to a Calabi-Yau threefold realized as a complete intersection in products of projective space.

One is faced with the following combinatorial problem:
\begin{problem}
Can we classify all configuration matrices (\ref{eqn:confmatr}) up to equivalence and redundancies?
\end{problem}
\noindent This represents one of the earliest big-data problems in the world of pure mathematics and physics.  It was undertaken in the late 1980s by Candelas, Lutken, Schimmrigk and others \cite{Candelas:1987kf}. Let us briefly survey the landscape of CICYs that were discovered:  

\begin{itemize}
\item There are \textbf{7890} CICYs corresponding to 7890 inequivalent configuration matrices.  The smallest matrix is $1 \times 1$ (corresponding to the quintic) and they reach a maximum of 12 rows or 15 columns.  

\item $q^{i}_{j} \in [0,5]$ for all $i,j$.  

\item There are 266 distinct Hodge pairs $(h^{1,1}, h^{2,1})$.

\item There are 70 distinct Euler characteristics $\chi \in [-200, 0]$.  

\item The transpose of a configuration matrix is again a configuration matrix.  

\item The 5 cyclic Calabi-Yau threefolds are the only ones with a single row.  In other words, there are only 5 complete intersection Calabi-Yau threefolds in a single projective space.  
\end{itemize} 

\begin{example}\label{ex:SchoenExample}
Consider the following configuration matrix
\begin{equation}\label{eqn:Schoen}
S=\left[
\begin{array}{cc}
1 & 1 \\
3 & 0 \\
0 & 3 \\
\end{array}
\right]
\end{equation}
From the conditions (\ref{eqn:CICYconds}), it is straightforward to check that $S$ corresponds to a compact Calabi-Yau threefold $X_{S}$ which is cut out of $\mathbb{P}^{1} \times \mathbb{P}^{2} \times \mathbb{P}^{2}$ by two equations of multi-degrees $(1,3,0)$ and $(1,0,3)$, respectively.  This is a CICY, which we call the \emph{Sch\"{o}en manifold}, after Chad Sch\"{o}en \cite{schoen}.  The two relevant Hodge numbers are $h^{2,1} = h^{1,1} =19$, and therefore, $\chi(X_{S}) =0$.  In the next section we will see that $X_{S}$ is also an elliptic fibration.  

The transpose of the matrix (\ref{eqn:Schoen}) 
\begin{equation}\label{}
TY=\left[
\begin{array}{ccc}
1 & 3 & 0 \\
1 & 0 & 3 \\
\end{array}
\right]
\end{equation}
of course also corresponds to a CICY, one called the Tian-Yau manifold $X_{TY}$.  The Hodge numbers are $h^{1,1}=14, h^{2,1}=23$ and therefore, $\chi(X_{TY}) =-18$.  The Tian-Yau manifold carries a free $G=\mathbb{Z}/3\mathbb{Z}$ action which preserves the Calabi-Yau structure.  As a result, the quotient $X_{TY}/G$ is a smooth compact Calabi-Yau threefold (though not a CICY) which has a special Euler characteristic $\chi = -6$, see \cite{CYLandscape,deeplearninglandscape,PLBpaper}.   At the time, this quotient was taken seriously as a candidate for the geometry of the universe!  Unfortunately, it has some problems in its matter content.    
\end{example}

\begin{figure*}[h!]
        \centering
        \begin{subfigure}[b]{0.375\textwidth}
            \centering
            \includegraphics[width=\textwidth]{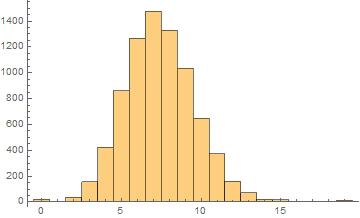}
            \caption[Network2]%
            {{\small $h^{1,1}$}}    
            \label{fig:mean and std of net14}
        \end{subfigure}
        \hfill
        \begin{subfigure}[b]{0.375\textwidth}  
            \centering 
            \includegraphics[width=\textwidth]{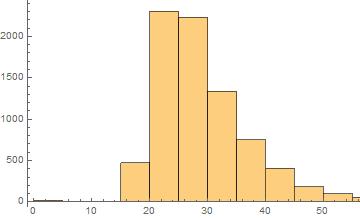}
            \caption[]%
            {{\small $h^{2,1}$}}    
            \label{fig:mean and std of net24}
        \end{subfigure}
        \vskip\baselineskip
        \begin{subfigure}[b]{0.375\textwidth}   
            \centering 
            \includegraphics[width=\textwidth]{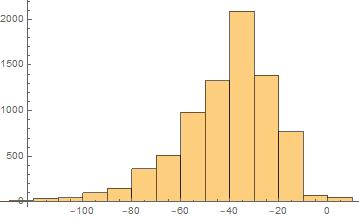}
            \caption[]%
            {{\small $\chi$}}    
            \label{fig:mean and std of net34}
        \end{subfigure}
        \quad
        \begin{subfigure}[b]{0.375\textwidth}   
            \centering 
            \includegraphics[width=\textwidth]{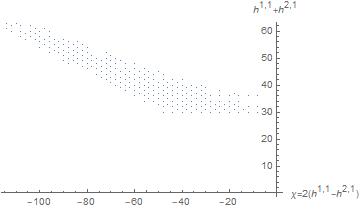}
            \caption[]%
            {{\small}}    
            \label{fig:mean and std of net44}
        \end{subfigure}
\caption{CICY topological data}
        \label{fig:TopDatPlot}
    \end{figure*}

In general, it is difficult to compute the Hodge numbers for the CICY dataset (in the above example, we gave them without proof).  We present this topological data, along with the Euler characteristic for CICYs in Figure \ref{fig:TopDatPlot}.  The Hodge numbers are presented as frequency plots.  Interestingly, the distribution of $h^{1,1}$ is somewhat Gaussian while $h^{2,1}$ is somewhat Poisson.

%

All CICYs have non-positive Euler characteristic.  One weak form of the mirror symmetry conjecture is that compact Calabi-Yau threefolds come in pairs with opposite Euler characteristics.  Therefore, if one put too much stock in the CICY dataset, they might wrongly convince themselves that all Calabi-Yau manifolds have negative Euler characteristic!  We clearly have to venture further in the landscape to encounter the mirror partners of the CICYs. 

\section{Elliptically Fibered Calabi-Yau Threefolds}

Elliptic curves are among the most beautiful objects in mathematics.  They provide a link between the fields of geometry, number theory, algebra, and even physics.  In fact, as we saw in Section \ref{sec:CYGinMP}, an elliptic curve is the unique Calabi-Yau manifold in dimension 1.  The notion of an \emph{elliptic fibration} should be thought of as elliptic curves moving in a family.  To understand this vague intuition, let us start with some basics.

Let $\Lambda \subset \mathbb{C}$ be a full-rank lattice.  Topologically, the quotient space $\mathbb{C}/\Lambda$ is a complex torus, or a Riemann surface of genus 1.  The following important proposition says that all such Riemann surfaces arise from cubic curves in the projective plane, i.e cubic plane curves.  

\begin{proposition}
Riemann surfaces of genus 1 are in bijection with smooth cubic hypersurfaces in $\mathbb{P}^{2}$, i.e. smooth vanishing loci of homogeneous degree 3 polynomials in 3 variables \cite{Griffiths}.   
\end{proposition}
\noindent By the degree-genus formula for plane curves, any smooth cubic hypersurface in $\mathbb{P}^{2}$ has genus 1.  Conversely, given a complex torus of the form $\mathbb{C}/\Lambda$, the Weierstrass $\wp$-function $\wp(\tau, z)$ associated to $\Lambda$ gives an embedding into $\mathbb{P}^{2}$.  And the differential equation satisfied by $\wp(\tau, z)$ implies that the image satisfies a cubic equation.

Consider the complex threefold $X \subset \mathbb{P}^{2} \times \mathbb{P}^{2}$ defined by the vanishing locus of the following bi-homogeneous degree (1,3) polynomial
\begin{equation}\label{eqn:ellfibEX}
a_{0}x_{0}^{3} + a_{1}x_{1}^{3} + a_{2}x_{2}^{3} = 0.  
\end{equation}
Here $(a_{0}:a_{1}:a_{2})$ are coordinates on the first factor of $\mathbb{P}^{2}$ and $(x_{0}:x_{1}:x_{2})$ are coordinates on the second.  Notice that for any point $(a_{0}:a_{1}:a_{2}) \in \mathbb{P}^{2}$ the above equation becomes a cubic in (the second) $\mathbb{P}^{2}$.  Therefore, the map $\pi : X \to \mathbb{P}^{2}$ defined by projection onto the first factor, is surjective and all fibers are cubics in $\mathbb{P}^{2}$.  This motivates the following definition.  

\begin{definition}
An \emph{elliptic fibration} is a morphism\footnote{Strictly speaking, we want the map $\pi$ to be flat and proper.  These are technical algebro-geometric conditions ensuring we have nice family of projective curves of arithmetic genus 1.} $\pi : X \to B$ between smooth algebraic varieties $X, B$ such that a generic fiber of $\pi$ is a smooth elliptic curve.  We call $X$ the total space and $B$ the base.  
\end{definition}
\noindent An \emph{elliptically fibered Calabi-Yau threefold}, is a Calabi-Yau threefold $X$ together with the structure of an elliptic fibration $\pi : X \to B$.  

One should think of an elliptic fibration $\pi : X \to B$ as a \emph{family} of elliptic curves parameterized by the base $B$.  However, over certain loci in the base, the elliptic curves can degenerate to singular curves.  In virtually all interesting fibrations in algebraic geometry, one has to allow for singular fibers.  For example, looking back to (\ref{eqn:ellfibEX}) the fiber above the point $(1:-1:0) \in \mathbb{P}^{2}$ is
\begin{equation*}
x_{0}^{3}-x_{1}^{3} = (x_{0}-x_{1})(x_{0}^{2} + x_{0}x_{1} + x_{1}^{2})
\end{equation*}
which is not a smooth cubic: it is the union of a line and a conic.  

\begin{example}\label{ex:RatlELL}
Let $Y \subset \mathbb{P}^{1} \times \mathbb{P}^{2}$ be the vanishing locus of the bi-homogeneous degree $(1,3)$ polynomial
\begin{equation*}
a_{0}f(x_{0}, x_{1}, x_{2}) + a_{1}g(x_{0}, x_{1}, x_{2}) = 0
\end{equation*}
where $f, g$ are generic homogeneous cubic polynomials.  Since for any point $(a_{0}, a_{1}) \in \mathbb{P}^{1}$, the above equation becomes a cubic in $\mathbb{P}^{2}$, the projection onto the first factor $\pi : Y \to \mathbb{P}^{1}$ defines an elliptic fibration called a \emph{rational elliptic surface}.  
\end{example}

\begin{example}
Recall from Example \ref{ex:SchoenExample}, the Sch\"{o}en manifold $X_{S} \subset \mathbb{P}^{1} \times \mathbb{P}^{2} \times \mathbb{P}^{2}$ is the vanishing locus of homogeneous polynomials of multi-degree $(1,3,0)$ and $(1,0,3)$ respectively.  Let $Y$ be a rational elliptic surface from Example \ref{ex:RatlELL}.  We can define a map 
\begin{equation*}
\pi : X_{S} \to Y \subset \mathbb{P}^{1} \times \mathbb{P}^{2}
\end{equation*}  
by projecting onto the vanishing locus of the multi-degree $(1,3,0)$ polynomial.  The fiber over a point in $\mathbb{P}^{1} \times \mathbb{P}^{2}$ is a cubic in $\mathbb{P}^{2}$ since we have to impose the second equation defining $X_{S}$.  Therefore, $X_{S}$ is an elliptically fibered Calabi-Yau threefold. 
\end{example}

The above example illustrates that there are CICYs which are also elliptically fibered Calabi-Yau threefolds.  See Figure \ref{fig:CYLandSum}, where ``S" denotes the Sch\"{o}en manifold.  According to \cite{CYLandscape,deeplearninglandscape,PLBpaper}, there is a common belief that ``most" Calabi-Yau threefolds are elliptically fibered.  It is an active area of research to determine precisely which Calabi-Yau threefolds are elliptically fibered.

\section{Additional Regions of the Compact Landscape} 

Unfortunately, there are many important classes of compact Calabi-Yau threefolds which we cannot discuss in detail here.  Most notably, the Calabi-Yau hypersurfaces in $4$-dimensional toric varieties.  This problem was undertaken in the late 1990s by Kreuzer-Skarke (KS), and resulted in one of the biggest datasets seen in pure mathematics.  For details on the KS dataset, we refer the reader to \cite{CYLandscape,deeplearninglandscape,PLBpaper}.  

\begin{figure}[h]
\centering
\includegraphics[width=95mm]{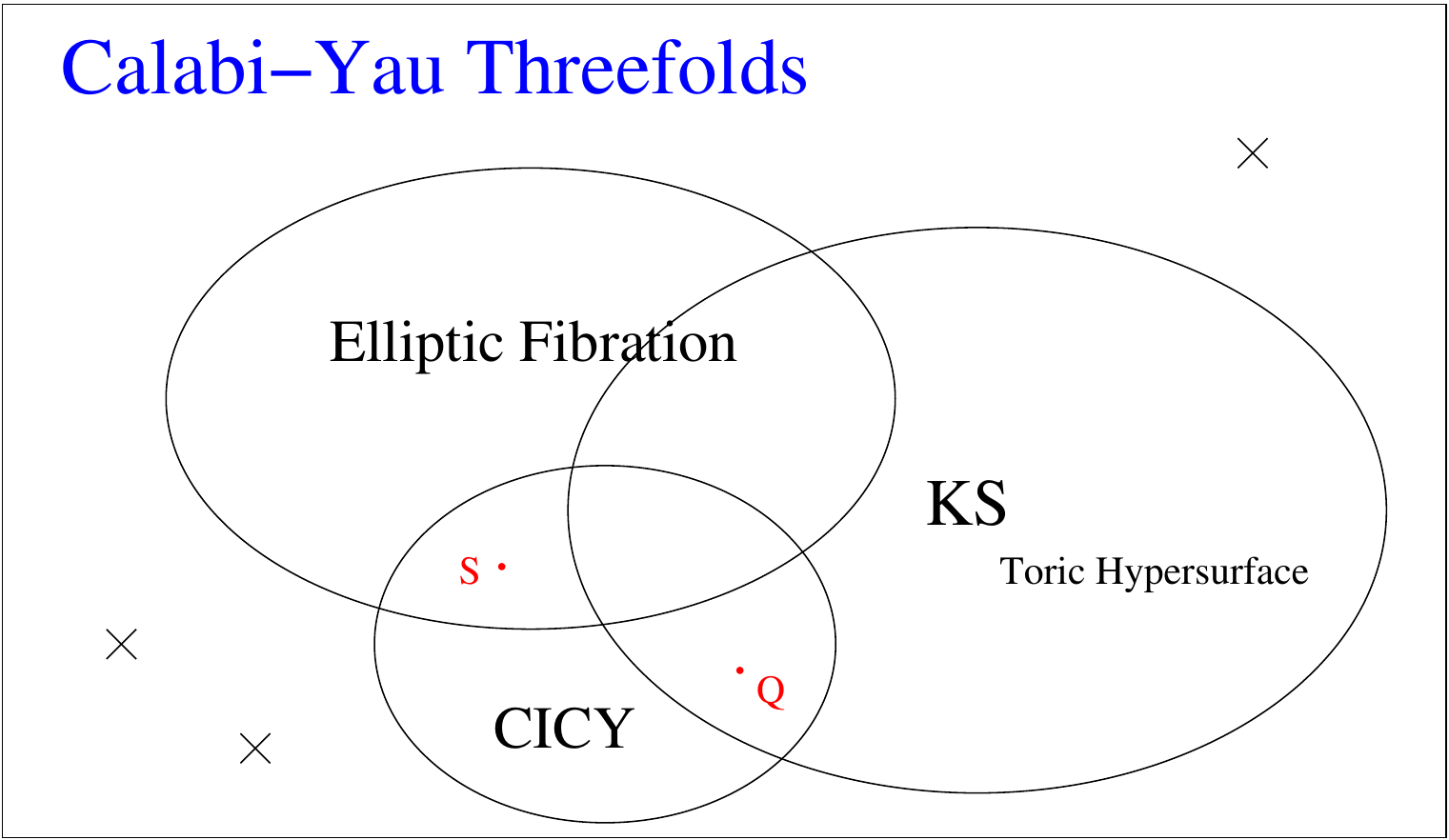} 
\caption{The compact Calabi-Yau threefold landscape}
\label{fig:CYLandSum}
\end{figure}

In Figure \ref{fig:CYLandSum} we summarize the portions of the Calabi-Yau landscape mentioned in this survey.  The point marked ``S" denotes the Sch\"{o}en manifold, which is both an elliptic fibration and a CICY.  The point marked ``Q" is the quintic, which is both a CICY as well as a hypersurface in a toric variety.  The points labelled ``$\times$" denote compact Calabi-Yau threefolds not falling into any of these groups.

\clearpage
\part{Non-compact Calabi-Yau Landscape}\label{part:NCCYL}

\section{String Theory Structures}

\subsection{D-branes}
D-branes occur in Type IIB Superstring theory as the Dirichlet boundary conditions of open strings. A D-brane is hence the hyperplane traced out by the allowed movement of the endpoint of an open string. The dimensionality of the D-brane in question defines the restriction on the directions the string endpoint can move in; such that a D$p$ brane only allows string endpoints to move in its ($p$+1)-dimensional world-volume. For example, a D0 brane is a spatial point moving through time, and fixes the endpoint of the string. Additionally, a D1 brane is a spatial line, forming a sheet as it is traced through time, and restricts the string endpoint to any position on this line for all time progression.
    \begin{figure}[h!]
        \centering
        \includegraphics[width=75mm]{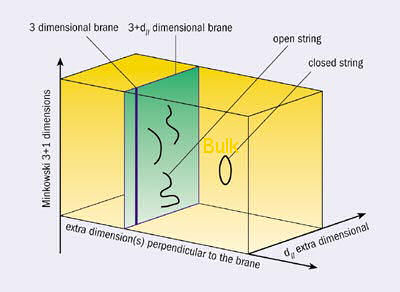} 
        \caption{A graphic representation of a D-brane \cite{brane_image}. The vertical axis gives full Minkowski space, $\mathbb{R}^{1,3}$, such that a vertical line is the D3 brane considered in Superstring theory. Further theories may use higher dimensional branes indicated by the vertical line's extension into a plane along the $d_\parallel$ axis. The remaining dimensions of the theory are extra, and only endpoints of open strings are restricted to the D-brane as shown.}
        \label{Dbrane}
    \end{figure}

The D-branes world-volumes support a tensor form of dimension ($p$+1), this can be integrated over the spatial dimensions to give a conserved charge, known as the Chan-Paton factor of the brane. The form in consideration connects the brane with a U(1)-bundle, such that enhanced gauge symmetry arises as the branes are stacked. In the stacking process, $N$ D-branes' world-volumes are overlaid in spacetime at an infinitesimal limit, and the total brane gauge group enhances via: $\text{U}(1)^N \mapsto \text{U}(N)$. Here the gauge connection on the branes generalises to a higher rank tensor as the string endpoints can be connected across multiple branes in the stack. This becomes important in defining the quiver representation, which is used in the following machine-learning analysis.
    
D-branes are important in the brane-world physical interpretation of Type II Superstring theory. In the 10-dimensional spacetime of the Type IIB superstrings, the endpoints are restricted to exist on a D3 brane, whose world-volume is the familiar $\mathbb{R}^{1,3}$ Minkowski space of general relativity and other theories. The remaining six dimensions form a non-compact Calabi-Yau space, such that $X^{10} = \mathbb{R}^{1,3} \otimes X^6$. The standard model exists on the D3 brane (or stack of $N$ D3 branes), and only interacts with the $X^6$ Calabi-Yau space via gravitation.
    
The simplest case of a non-compact Calabi-Yau 3-fold is $\mathbb{C}^3$, which is trivially Ricci-flat. Beyond that Orbifolds are a natural candidate. Orbifolds are formed from action of a discrete group quotient on a manifold. These manifolds are discussed further in Appendix \ref{toric_varieties} \cite{CYLandscape,deeplearninglandscape,PLBpaper}. 
    
\subsection{Quivers}
A Quiver, $\mathcal{Q}$, is a multi-digraph, such that its set of nodes and arrows have finite cardinalities $N_0$ and $N_1$ respectively. The quiver represents a gauge theory, where each node has an associated U($N_i$) gauge group. The product of all node gauge groups give the full gauge group of the theory. Each arrow is associated with a field, $X_{ij}$, in the bi-fundamental representation of the gauge groups associated with the nodes connected to the arrow. The fields transform according to the Young tableaux $(\square,\overline{\square})$ for the nodes groups. The superpotential, $W$, of the theory the quiver represents leads to a set of polynomials, $\{\partial_{X_{ij}}W=0\}$, which physically give the vacuum state of the theory.
        
Importantly, the representation variety of the quiver is the Vacuum Moduli Space of the gauge theory. A quiver's representation variety is the gauge invariant quotient of the quiver's representations, with relations from the superpotential, and quotiented by a product group of complex General Linear transformations. Geometric invariant theory (GIT) is generally used to construct moduli spaces by considering the quotients of groups on algebraic varieties.

\begin{wrapfigure}[12]{r}{0.5\textwidth}
    \begin{center}
    \includegraphics[width=40mm]{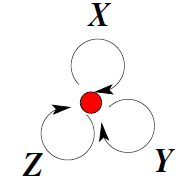} 
    \caption{The quiver for $\mathcal{N}=4$ Super Yang-Mills theory, with three adjoint fields: X, Y, Z \cite{CYLandscape,deeplearninglandscape,PLBpaper}.}
    \label{quiver}
    \end{center}
\end{wrapfigure}
This representation variety is an affine variety, such that the coefficients of the zero-locus of the variety's polynomial set generates the corresponding prime ideal. Conversely, the Vacuum Moduli Space of a gauge theory is a geometric space with a vacuum state of the gauge theory associated to each point in the space. This moduli space often forms a manifold known as the vacuum manifold of the theory.

The space of quivers and superpotentials, $(\mathcal{Q},W)$, produces a space of representation varieties, which hence give all the Vacuum Moduli Spaces of the gauge theory in question. Each of the Vacuum Moduli Spaces of a supersymmetric gauge theory is a non-compact Calabi-Yau manifold, and hence this is how the non-compact Calabi-Yau landscape naturally arises in Superstring theory. A simple example of a quiver is the ``clover'', which represents the famous $\mathcal{N}=4$ Super Yang-Mills theory, shown in figure \ref{quiver}. The superpotential for this example is $W = \text{Tr}\big([X,Y]Z\big)$ which leads to the simplest Vacuum Moduli Space case of $\mathbb{C}^3$ \cite{CYLandscape,deeplearninglandscape,PLBpaper}.

\subsection{An Orbifold Example: $\mathbb{C}^3/\mathbb{Z}_3$}
\begin{figure}[h!]
    \centering
    \includegraphics[width=40mm]{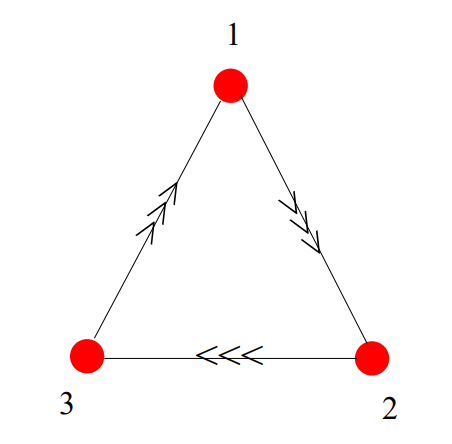} 
    \caption{The U$(1)^3$ quiver with 9 fields denoted by the 3 sets of 3 arrows \cite{Hauenstein:2012xs}.}
    \label{C3Z3_quiver}
\end{figure}

Here we consider a typical example of quiver gauge theory used commonly in association with AdS/CFT correspondence, as it is the worldvolume theory of a D3 brane in the bulk spacetime. This non-compact Calabi-Yau manifold examined is given by the toric variety: $\mathbb{C}^3/\mathbb{Z}_3$. This quotient structure makes the manifold an orbifold; where the algebraic geometry structure is explained further in Appendix \ref{toric_varieties}. The U$(1)^3$ quiver in question is shown in figure \ref{C3Z3_quiver} and shows 9 fields in the theory.

Since 3 fields exist on each of the 3 edges, there are correspondingly $3^3 = 27$ gauge invariant operators possible, associated with all the closed cycles in the quiver. In this theory, each of the gauge invariant operator terms appear in the superpotential as products of the fields in the corresponding cycle, giving
\begin{equation}
    W = \sum^3_{\alpha,\beta,\gamma = 1} \varepsilon_{\alpha\beta\gamma} X^\alpha_{12} X^\beta_{23} X^\gamma_{31}\,,
\end{equation}
for the totally antisymmetric rank 3 tensor $\varepsilon_{\alpha\beta\gamma}$\,, where the Greek indices run $1\mapsto 3$ for each of the 3 arrows between each pair of nodes. Each field has subscripts to denote the nodes it is in representations of. The are also 9 F-term equations of motion from the superpotential, which are
\begin{equation}
    0 = \sum^3_{\beta,\gamma =1} \varepsilon_{\alpha\beta\gamma} X^\beta_{23}X^\gamma_{31} = \sum^3_{\alpha,\gamma =1} \varepsilon_{\alpha\beta\gamma} X^\alpha_{12}X^\gamma_{31} = \sum^3_{\alpha,\beta =1} \varepsilon_{\alpha\beta\gamma} X^\alpha_{12}X^\beta_{23}\,,
\end{equation}
where each term is 3 equations for each value of the uncontracted index. These equations arise under the action of $0 = \partial_{X}W$ for each of the fields, $X$. 

The 27 gauge invariant operators are redefined as dimensions of $\mathbb{C}^{27}$, denoted $y_{\alpha\beta\gamma}$. Elimination with the F term equations via low degree polynomial interpolation \cite{Hauenstein:2012xs} leads to a system of 17 linear equations, and 27 quadratic equations. Further elimination via trivial substitution with the 17 linear equations reduces the system to 27 equations in 10 variables, thus giving the $\mathbb{C}^{10}$ space. These equations are recognised as the standard Veronese embedding: $\mathbb{P}^2 \hookrightarrow \mathbb{P}^9$ which can be affinised into a $\mathbb{C}^3$ embedding in the $\mathbb{C}^{10}$ found above. This embedding corresponds to their existing exactly 10 degree 3 monomials in 3 variables, such that each one then corresponds to a dimension in $\mathbb{C}^{10}$. These equations then give the degree 9 irreducible variety which defines the 3 dimensional orbifold $\mathbb{C}^3/\mathbb{Z}_3$ \cite{Gray:2006jb}.

\begin{figure}[h!]
    \centering
    \includegraphics[width=40mm]{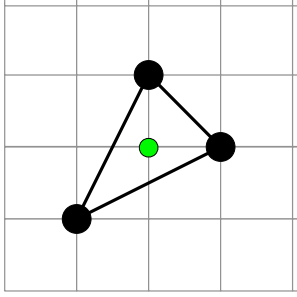} 
    \caption{The toric diagram dual for the $\mathbb{C}^3/\mathbb{Z}_3$ orbifold \cite{CYLandscape, deeplearninglandscape, PLBpaper}. The origin is denoted in the diagram centre, and the toric diagram can be retrieved as this diagram's dual.}
    \label{C3Z3_toric_dia}
\end{figure}

The exponents of the 10 degree 3 monomials in 3 variables give vectors in the fan of the toric variety definition (noting that the orbifold being abelian makes it also toric). Taking the rays of this fan gives three coplanar vectors, which in the plane correspond to points which in turn define the toric diagram. These are $\{(1,0),(0,1),(-1,-1)\}$,which are plotted in figure \ref{C3Z3_toric_dia}, the dual of this diagram then gives the orbifold's toric diagram \cite{CYLandscape, deeplearninglandscape, PLBpaper, Hauenstein:2012xs}.

The corresponding brane tiling and dessin d'enfant can then be formed from the toric diagram; these objects are addressed in section \ref{AGV} \cite{Franco:2017jeo}.

\subsection{McKay Correspondence}

McKay correspondence concerns a discrete finite subgroup $G \subset \text{SU}(2)$. Firstly taking the tensor product between the defining 2 complex dimensional rep of $G$ and each irrep of $G$, and then taking the irrep decomposition of this tensor product makes the correspondence manifest. Whereby each decomposition coefficient is the square of the adjacency matrix for each of the simply-laced Dynkin diagrams.

Dynkin diagrams represent the root system of the gauge group's Lie algebra. To be simply-laced means there is only one edge between each node, which represents a restriction on the angles between the fundamental roots. Specifically, the simply-laced Dynkin diagrams are: $A_n$, $D_n$, $E_n$ where the first two are series of diagrams for $n \in \mathbb{Z}^+$, whilst the $E_n$ refers to three of the exceptional Lie algebras. The Dynkin diagrams in question are affine-extended, which is canonically achieved by central extension of the original Lie algebra. This amounts to introducing an additional imaginary root, which increases the dimensionality of the root system. These are hence represented with an additional node, and denoted: $\tilde{A}_n$, $\tilde{D}_n$, and  $\tilde{E}_n$ respectively. In the special case of simply-laced, the Dynkin diagrams correspond exactly to their Coxeter diagrams, which represent Coxeter groups, defined by reflection symmetries.
    
This is relevant because the representation variety of the affine Dynkin diagrams formulated as quivers are Calabi-Yau 2-folds (a.k.a. K3-surfaces). We can then produce orbifolds from these described by McKay quivers such that they have the form $\mathbb{C}\times(\mathbb{C}^2/G)$. These orbifolds are hence also candidates for the extra dimensions in Superstring theory. However where $\mathbb{C}^3$ leads to $\mathcal{N}=4$ Super Yang-Mills theory on the D3 brane, these orbifolds produce $\mathcal{N}=2$ supersymmetric QFTs.
    
When taking quotients to produce the orbifolds in question, relations between the invariants of the orbifold give rise to algebraic singularities. In $\mathbb{C}^2$ these are the du Val singularities. Smoothing out these singularities through desingularisation requires the resolution map between the canonical bundle and canonical sheaf to be crepant. Meaning that no discrepancy divisor is needed with the resolution map. When this crepant resolution map is established, metrics and other physically relevant measures can be written explicitly for some special cases.
    
These crepant resolutions are key in generalising the quotient process to act on Calabi-Yau 3-folds (as $\mathbb{C}^3/G$); introducing further orbifolds into the Calabi-Yau spectrum. However in this case the manifolds are related to one another by mirror symmetry and in particular flop transitions. These orbifolds correspond to $\mathcal{N} = 1$ super-conformal gauge theories, hence extending also the practical applications of studying the Calabi-Yau landscape with respect to examining topical theories in physics.
    
The quotient product, and crepant resolution methods extend the landscape of non-compact Calabi-Yau manifolds from only $\mathbb{C}^3$ to also include a plethora of orbifolds. Physicists interpret the manifold landscape as representation varieties of quivers, which indicate the equivalent gauge-field theories \cite{CYLandscape,deeplearninglandscape,PLBpaper}.

\section{Algebraic Geometry Viewpoint}\label{AGV}

\subsection{Brane Tilings}
\begin{figure}[h!]
        \centering
        \includegraphics[width=150mm]{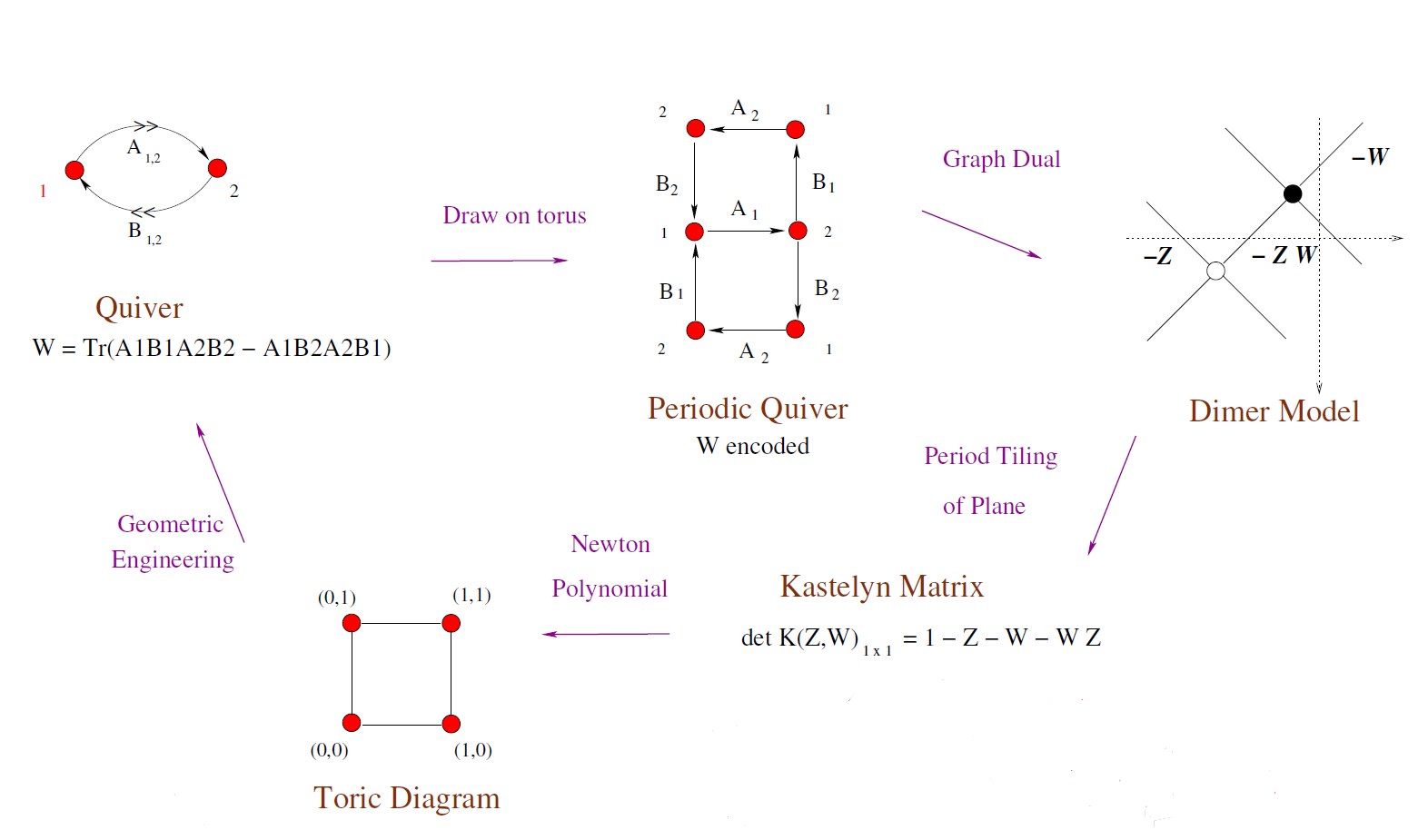} 
        \caption{A pictorial representation of the process that links the quiver and superpotential ($\mathcal{Q}$,W) to the Toric diagram of the equivalent non-compact Calabi-Yau manifold \cite{CYLandscape,deeplearninglandscape,PLBpaper}. This specific example is for the conifold considered previously.}
        \label{QtoD}
    \end{figure}

The method to connect the quivers of a gauge theory, with the toric diagram (see Appendix \ref{toric_varieties}) of the relevant Calabi-Yau that makes up the remaining dimensions in the full 10d superstring spacetime, exists for both directions \cite{Hanany:2005ve, Franco:2005sm}. Deriving the toric diagram from the quiver is more straightforward and follows the clockwise process depicted in figure \ref{QtoD}.
    
The converse, ``geometric engineering'', toric diagram to quiver process was originally computationally demanding, with exponential time complexity. This process was streamlined by introducing the concept of brane tiling. This brane tiling concept was derived from noticing a consistent relation between the number of nodes, edges, and superpotential terms, $(\mathcal{N}_0,\mathcal{N}_1,\mathcal{N}_2)$ respectively, 
    \begin{equation}\label{euler}
        \mathcal{N}_0 - \mathcal{N}_1 + \mathcal{N}_2 = 0\,.
    \end{equation}
This applied for all quivers whose representation variety was a toric variety (as for those considered in string theory). The relation \ref{euler} was associated to the Euler characteristic for a torus, and this allowed the quiver and superpotential to be encoded as a bipartite graph tiling on a (genus, $g=1$) torus. The connection of brane tilings to quivers follows a simple algorithm. Whilst mapping from the toric diagrams to the brane tilings is epimorphism; with the orbit of tilings which are mapped to by the same toric diagram related by Seiberg duality \cite{Yamazaki}.
    
\begin{figure}[h!]
    \centering
    \includegraphics[width=60mm]{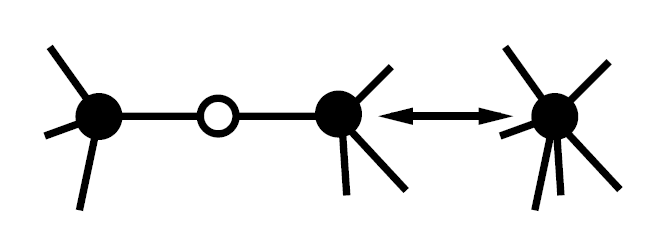} 
    \caption{The contraction of part of a brane tiling \cite{Yamazaki}, corresponding to integrating out a massive field to relate two quiver gauge field theories via Seiberg duality.}
    \label{seiberg}
\end{figure}

Seiberg duality relates an ``electric'' and a ``magnetic'' theory, stating that under RG flow they both approach the same IR fixed point. Therefore they represent the same theory at lower energy densities. In our context it represents the relation between two quiver gauge theories, where some additional fields are integrated out/introduced, which graphically corresponds to contracting/expanding parts of the brane tilings. This concept is exemplified in figure \ref{seiberg}.
    
More mathematically, the Seiberg duality process corresponds to cluster mutation of the mathematical graph-theoretic quiver objects. Through a series of steps of reorienting and reassigning arrows associated with a node in the quiver, and adjusting the gauge group size by the number of fields, a different (dual) quiver is formed \cite{cluster}. This cluster mutation process is a generalisation of the Seiberg duality. Multiple actions of the cluster mutation for different nodes creates ``mutation classes'' of quivers. Their equivalent brane tilings are connected by a process known as urban renewal, again a mathematical generalisation of the integrating out/introduction of fields in the physical application of Seiberg duality. These dual quivers are related, where their tilings correspond to the same toric diagram under the epimorphism previously mentioned. Tilings are an important step in the geometric engineering process.
    
The quiver duality concept may also be thought of as monodromy of wrapped 3-cycles in the dual theory via another duality known as mirror symmetry. Mirror symmetry connects mirror dual Calabi-Yau manifolds in different superstring theories, where they lead to the same resulting physics. In this case the D3 brane on one Calabi-Yau 3-fold is mirror dual to a D6 brane with 3 dimensions identified (3-cycle wrapping) on the dual Calabi-Yau 3-fold. This concept has been shown to be practical in Topological string theory where the mirror symmetry concept has been mathematically well defined \cite{witten_mirror}.
    
Mirror symmetry allows calculation of certain complicated invariants by performing easier calculations in the dual theory. A key example is Gromov-Witten invariants, which arise in symplectic geometry which also satisfies the 'almost complex' structure requirements. The almost complex structure is a looser condition than K\"ahler geometry in that only the tangent space is required to be smooth linear complex, and not necessarily the underlying space. These invariants are calculated from pseudoholomorphic curves which are the symplectic equivalent of distances in Riemannian geometry. More general quantities are usually expressed in terms of the Gromov-Witten invariants, which are difficult to compute, but can be reduced to simpler integrals in the mirror dual theory \cite{mirror,Yang_mirror}.

\subsection{Dessin d'Enfants}
Bipartite tilings are the algebraic geometry equivalent of Grothendieck's Dessin d'Enfants from number theory. This interpretation can be useful for categorising the tilings, and hence the quiver gauge theories. Mathematically the dessins are interpreted using Belyi maps, $\beta$, which map from a smooth compact Riemann surface (described as a hyperelliptic curve of complex numbers), $\Sigma$, to projective space, $\mathbb{P}^1$ such that \cite{Jejjala:2010vb}
    \begin{equation}
        \beta: \Sigma \longmapsto \mathbb{P}^1\,.
    \end{equation}
A dessin is then formed from the preimage of a Belyi map which has three ramification points; where a ramification point is an element of $\Sigma$ where the local Taylor expansion of $\beta$ starts at order $\geq 2$ and corresponds to degeneration of the map. Under the SL(2,$\mathbb{C}$) symmetry of $\mathbb{P}^1$, the three ramification points are transformed to $(0,1,\infty)$, and the dessin is formed by associating the preimages of 0 to black nodes, preimages of 1 to white nodes, and preimages of the (0,1) interval to edges. Therefore the dessin is a bipartite graph drawn on the Riemann surface $\Sigma$, such that
    \begin{equation}
        \beta^{-1}(0) \rightarrow \bullet\,, \quad \beta^{-1}(1) \rightarrow \circ\,, \quad \beta^{-1}(0,1) \rightarrow -\,.
    \end{equation}
The dessins can be categorised by their passports, which is the collection of the ramification data, represented
    \begin{equation}
        \big[r_0(1),r_0(2),...,r_0(B) |r_1(1),r_1(2),...,r_1(W) |r_\infty(1),r_\infty(2),...,r_\infty(I)\big]\,,
    \end{equation}
such that $r_i(j)$ is the ramification value (order of the leading term in Taylor expansion) of the $j$th preimage of value $i$ in the image of the Belyi map. The total number of preimage points are $(B,W,I)$ for the ramification points $(0,1,\infty)$ respectively. Note also here that the Riemann-Hurwitz formula sets $B=W$ for the genus 1 torus we are working on. The actual value of each ramification point then gives the valency of each node in the dessin.
    
The passport doesn't identify the dessins exactly, a more effective way of representing the dessins independently is combinatorically as permutation triples. Permutation triples encode the dessin information by creating elements of the symmetric group which are the products of all cycles containing either the white nodes, $\sigma_W$, or the black nodes, $\sigma_B$. An additional object, $\sigma_\infty$, in the symmetric group is defined also, such that: 
    \begin{equation}
        \sigma_W \cdot \sigma_B \cdot \sigma_\infty = \mathbb{1}_d\,,
    \end{equation} 
for $\mathbb{1}_d$ the identity element of the symmetric group, $S_d$, of the $d$ edges in the dessin. The group elements $\sigma_\infty$ are then associated to cycles about faces of the dessin under this symmetric group \cite{Yang_dessin}.
     
In supersymmetric QFTs, R-symmetry connects the fields in the theory via their R-charge of the supersymmetric representations. For the tiling, each edge has an R-charge from the field it represents in the superpotential interpretation of the tiling. Under the symmetry, these charges must satisfy
    \begin{equation}\label{Rcharge}
        \sum_{i \in \mathfrak{E}_n}R_i = 2\,,\qquad \sum_{i \in \mathfrak{E}_f}\big(1-R_i\big) = 2\,,
    \end{equation}
for $\mathfrak{E}_n$ the edges bounding any node in question, and $\mathfrak{E}_f$ the edges bounding any face in question. These relations in terms of the tiling are equivalent to Euler's relation as in equation \ref{euler}.
    
Isoradial embedding is a method for constructing the tiling which automatically satisfies these required conditions on the R-charges. Nodes are organised on the circumferences of intersecting tessellated circles such that the angles subtended by triangles formed with adjacent nodes and the circle's centres satisfy $\theta_i = \pi R_i / 2$. This causes the conditions in \ref{Rcharge} to translate to basic geometric conditions on total angle around a point and total interior angle of a polygon respectively. These conditions fix the nodes' positions up to rotation about the circles, this is then fixed by performing $a$-maximisation of the function 
    \begin{equation}
        a(R_i) = \sum_{i \in \mathfrak{E}} (R_i - 1)^3\,, 
    \end{equation}
for $\mathfrak{E}$ the set of all edges in the tiling. Maximising this equation over the $R_i$ partition is equivalent to minimising the conic base volume.
    
    
\section{Non-compact Calabi-Yau Summary}

The non-compact Calabi-Yau landscape makes itself of manifest importance in superstring theory through the interpretation of quiver gauge theories. Within this, the manifolds make up the additional dimensional space in the theories' brane world interpretation. Using generalizations of McKay Correspondence, the association from the manifolds to the quivers is made through their representation varieties. Orbifolds can then be introduced into the landscape using group quotients and crepant resolution.
    
Alternatively the manifolds may be considered more algebraically as toric varieties, generally defined using a fan structure on a lattice. The general toric variety construction allows formation of more Calabi-Yau manifolds, including the conifold. From this interpretation toric diagrams can be formed from the varieties which aid in manifold classification, especially in the context of brane tilings.
    
Brane tilings are a useful geometric interpretation of the quiver and superpotential properties; and particularly streamline the process of calculating the physical theories represented by toric diagrams. Seiberg duality and mirror symmetry also become important concepts in this consideration of forming physical quiver gauge theories.
    
\begin{figure}[h!]
        \centering
        \includegraphics[width=120mm]{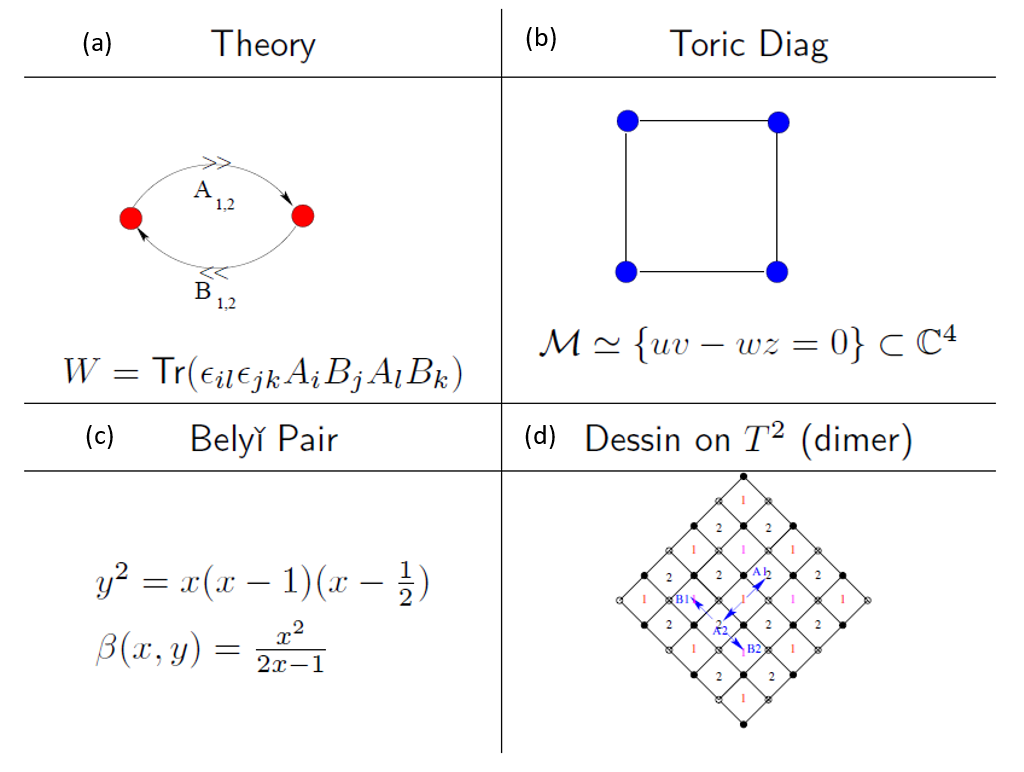} 
        \caption{The conifold Calabi-Yau manifold interpreted in terms of: (a) its underlying physical theory in terms of quiver and superpotential; (b) the representation variety's toric diagram (note it is the equivalent dual diagram that is shown); (c) the Belyi pair used to encode the dessin tiling structure; and (d) the brane tiling on a torus \cite{CYLandscape,deeplearninglandscape,PLBpaper}.}
        \label{conifold_full}
    \end{figure}

Finally these tilings can be considered in parallel to dessin d'enfants from number theory. This interpretation in terms of Belyi maps and their ramifications offers some explanation of structure associated with the underlying theories' supersymmetry. This interconnection between these interpretations of the Calabi-Yau manifolds is well depicted in the example in figure \ref{conifold_full} for the conifold.
    
Points of interest for further investigation include the interpretation of the Seiberg duality in terms of the dessin structure; and how the use of dessins may relate the absolute Galois group (important in the theory of dessins) into the physical theories of the tilings. Beyond these, the parallels between the physical, algebraic geometry, and number theoretic structures offers many sources for inspiration.

\clearpage
\part{Machine-Learning the Landscape}\label{machinelearning}

As we have seen above, different areas in mathematics, including algebraic geometry, representation theory and even number theory, have appeared in our study of CY manifolds in theoretical physics. As the extra six dimensions are believed to be ``wrapped'' as a CY 3-fold under string compactification, people began to search for the possible CY$_3$'s, and have so far collected a gigantic list of CY$_3$'s from reflexive polytopes, estimated at order $10^{10}$ \cite{deeplearninglandscape,PLBpaper}. Furthermore, the number of string vacua in the landscape is astonishingly of order $10^{500}$ for type \uppercase\expandafter{\romannumeral2}B theory\footnote{For F-theory, the number of flux vacua arisen from elliptic fourfolds even rockets to at least $10^{272000}$.} \cite{Ftheorylandscape}. Thus, the power of computers and algorithms is urgent for this interdisciplinary research. A different version of ``WWJD" has now been raised: what would $\mathtt{Jython}$ do?

\section{Performance Measures: Hypersurfaces in $W\mathbb{P}^4$}
\label{warmupWP4}

In Appendix \ref{introduction}, machine learning is briefly introduced. Whatever approach the machine adopts for the learning, we always need to know how well it performs. Let us quantify its performance using the following example. Recall the weighted complex projective space $W\mathbb{P}^4$ in (\ref{weightedproj})\footnote{In terms of the notation in (\ref{weightedproj}), this is $\mathbb{CP}^{(a_0,a_1,a_2,a_3)}$. For brevity, we will henceforth denote it as $W\mathbb{P}^4$.}. Our input for each hypersurface in $W\mathbb{P}^4$ is a 5-vector of co-prime positive integers which determines the space. Let us consider a simple query of whether the Hodge number $h^{2,1}>50$. Geometrically, we are searching for CY$_3$'s with a relatively large number of complex deformations. Our data $D$ consists of 7555 5-vectors, $x_i$, each resulting in a binary output, $y_i$. For example, $(\{1,1,1,1,1\}\rightarrow1)\in\{(x_i\rightarrow y_i)\}=D$ as $h^{2,1}=101>50$ in this case. On the other hand, we have $(\{2,2,3,3,5\}\rightarrow0)$ since $h^{2,1}=43<50$ here. In \cite{landauginzburg}, the Hodge numbers are computed using Landau-Ginzburg method. However, such procedure would take hours, and this is just a very simple query.

Now that our data $D$ is fully known, we can then split our data into a training set $T$ and a validation set $V$, viz, $D=T\bigsqcup V$. We can then establish a machine-learning algorithm so as to check how well it performs. This procedure is known as the \emph{cross validation}. To quantify the accuracy, we make the following definitions.
\begin{definition}
Let $V=\{(x_i\rightarrow y_i)\}$ where $y_i$ is the actual correct output for input $x_i$, and let $y_i^{\text{pred}}$ be the output predicted by the machine-learning model on $x_i$ with $i$ running from 1 to $N$. Then the \emph{precision} $p$ is the percentage that $y_i^{\text{pred}}$ agrees with $y_i$:
\begin{equation}
    p:=\frac{1}{N}|\{y_i^{\text{pred}}=y_i\}|\in[0,1].
\end{equation}
\end{definition}
\begin{definition}
Consider $y_i$ and $y_i^{\text{pred}}$ as vectors $\bm{y}$ and $\bm{y}^{\text{pred}}$ respectively. Then the \emph{cosine distance} is
\begin{equation}
    d_C:=\frac{\bm{y}\cdot\bm{y}^{\text{pred}}}{|\bm{y}||\bm{y}^{\text{pred}}|}\in[-1,1].
\end{equation}
If the cosine angle between the two vectors is 1, we have a complete agreement. If $d_C$ is -1, then it is the worst fit. If $d_C=0$, then it is a random correlation.
\end{definition}

Now we take 2000 samples (out of 7555) from $D$, which is approximately 25\%, to be our training data. Then we establish our MLP and test it using the remaining data. The detailed $\mathtt{Python}$ code can be found in \cite{CYLandscape,deeplearninglandscape,PLBpaper}. It turns out that there are only 375 errors in our experiment, which gives $p=(5555-375)/5555\simeq93.25\%$, and the cosine distance $d_C$ is 0.91. This is a quite impressive result with such a high accuracy. Remarkably, the running time is less than one minute on an ordinary laptop\footnote{This can also be done using $\mathtt{Mathematica}$ with a high accuracy as well. In particular, $\mathtt{Mathematica}$ has machine learning built into its core operating system from version 11.2, and now $\mathtt{Mathematica}$ 12 has been released with detailed documentation on machine learning.}!

To make sure that our machine-learning makes satisfying predictions, we need to introduce Matthews correlation coefficient (MCC). Firstly, we have:
\begin{definition}
Let \{($x_i\rightarrow y_i$)\} be categorical data, where $y_i\in\{1,2,\dots,k\}$ takes value in $k$ categories. Then the \emph{confusion matrix} is a $k\times k$ matrix where 1 is added to the ($ab$)$^{\text{th}}$ entry if the actual value of $y$ is $a$ while the predicted $y^{\text{pred}}$ is $b$.
\end{definition}
As a result, we want the confusion matrix to be diagonal ideally. In our binary case, the confusion matrix is $2\times2$, and we have this table:
\begin{equation}
    \begin{tabular}{|c|c|c|}
    \hline
      \diagbox{Predicted}{Actual} & True (1) & False (0) \\
    \hline
       True (1) & True Positive ($tp$) & False Positive ($fp$) \\
    \hline
       False (0) & False Negative ($fn$) & True Negative ($tn$) \\
    \hline
    \end{tabular}.
\end{equation}
Then we can define:
\begin{definition}
For binary classifications, the Matthews correlation coefficient is the square root of the normalized $\chi$-squared, that is,
\begin{equation}
    \phi:=\sqrt{\frac{\chi^2}{N}}=\frac{tp\cdot tn-fp\cdot fn}{\sqrt{(tp+fp)(tp+fn)(tn+fp)(tn+fn)}}\in[-1,1].
\end{equation}
\end{definition}
Such definition can also be generalized to $k\times k$ confusion matrices \cite{matthews,matthews1}. If the MCC returns 1, then we have a perfect prediction. If MCC is -1, then our fit is a complete disagreement. If $\phi=0$, then it is a random prediction. It is crucial to notice that other measures such as $p$ and $d_C$, unlike MCC, are not useful when the sizes of two classes differ too much, i.e., when we have \emph{imbalanced} data. For example, if there is only 0.1\% of the data to be classified as true. Then our algorithm would naively train a model predicting false for any input. Nevertheless, the accuracy $p$ would still reach 99.9\%\footnote{There are also other measures (especially required for imbalanced data) such as F-score. However, MCC is the most informative one as it includes all the four categories in confusion matrices \cite{matthews2}.}. In our case of hypersurfaces in $W\mathbb{P}^4$, we have $\phi=0.84$, which gives a quite nice prediction.

Now one may wonder how well our NN will behave when we change the number of samples in the training data. This can be analyzed via \emph{learning curves}:
\begin{definition}
Let $D=\{(x_i\rightarrow y_i)\}$ have $N$ data-points. We choose cross validation by taking $\gamma N$ data-points randomly as training data $T$, with some $\gamma\in(0,1]$. Then the remaining $(1-\gamma)N$ data-points form the validation data $V$. The performance of the machine-learning algorithm, upon training on $T$ and validated on $V$, is a function $L(\gamma)$ measured by any goodness of fit as aforementioned. The learning curve is then the plot of $L(\gamma)$ against $\gamma$.
\end{definition}
In practice, $\gamma$ is chosen discretely. Moreover, for each $\gamma$, we repeat random samples $\gamma N$ a number of times for statistical stability, so there are error bars associated to our points on the curve.

\begin{figure}[h]
    \centering
    \includegraphics[width=12.5cm]{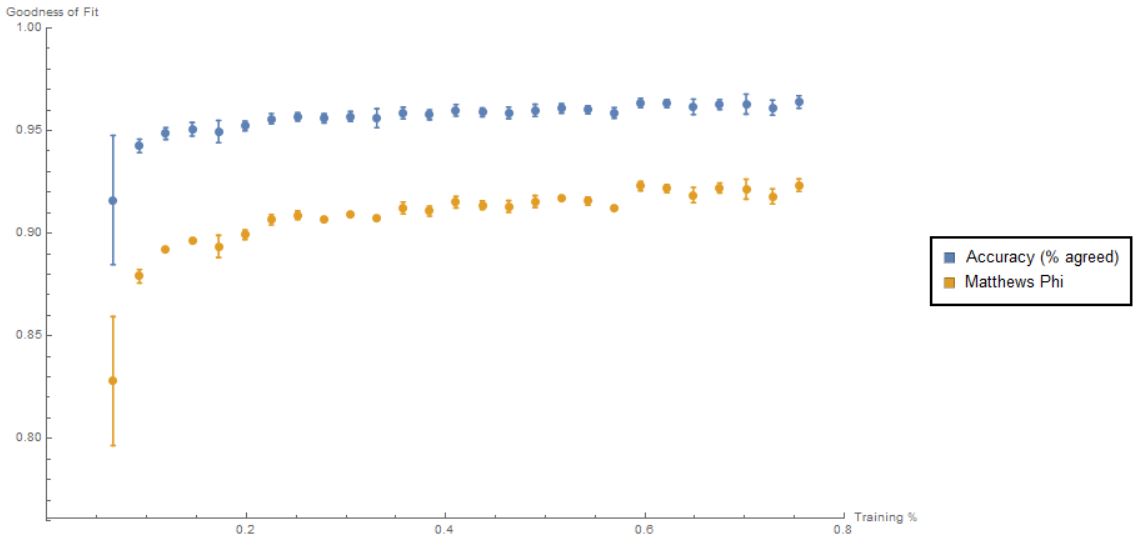}
    \caption{The learning curves for machine learning whether $h^{2,1}>50$ for a hypersurface in $W\mathbb{P}^4$. We repeat cross validation 10 times at each incremental interval of 5\%.}
    \label{wp4}
\end{figure}

The learning curve of our hypersurfaces in $W\mathbb{P}^4$ case is depicted in Fig. \ref{wp4}.
As we can see, there is a large error for training data less than 10\% as the NN has not seen enough data for valid predictions. However, from 20\%, our predictions become really well-behaved. The curve then ascends steadily as the (both) measures are approaching to 1.

\section{Learning CICYs}
\label{learningCICY}

\begin{figure}[h]
	\centering
	\begin{minipage}[b]{.45\textwidth}
		\centering
		\tiny
		\(
		\setcounter{MaxMatrixCols}{20}
		\begin{pmatrix}
		1&1  &0  &0  &0  &0  &0  &0  &0  &0  &0  &0 \\ 
		1&0  &1  &0  &0  &0  &0  &0  &0  &0  &0  &0 \\ 
		0&0  &0  &1  &0  &1  &0  &0  &0  &0  &0  &0 \\ 
		0&0  &0  &0  &1  &0  &1  &0  &0  &0  &0  &0 \\ 
		0&0  &0  &0  &0  &0  &2  &0  &0  &0  &0  &0 \\ 
		0&1  &1  &0  &0  &0  &0  &1  &0  &0  &0  &0 \\ 
		1&0  &0  &0  &0  &1  &1  &0  &0  &0  &0  &0 \\ 
		0&0  &0  &1  &1  &0  &0  &1  &0  &0  &0  &0 \\ 
		0&0  &0  &0  &0  &0  &0  &0  &0  &0  &0  &0 \\
		0&0  &0  &0  &0  &0  &0  &0  &0  &0  &0  &0 \\ 
		0&0  &0  &0  &0  &0  &0  &0  &0  &0  &0  &0 \\ 
		0&0  &0  &0  &0  &0  &0  &0  &0  &0  &0  &0 \\ 
		0&0  &0  &0  &0  &0  &0  &0  &0  &0  &0  &0 \\ 
		0&0  &0  &0  &0  &0  &0  &0  &0  &0  &0  &0 \\ 
		0&0  &0  &0  &0  &0  &0  &0  &0  &0  &0  &0 \\
		\end{pmatrix}
		\)
		\caption{The matrix representing the CICY\\ in $(\mathbb{P}^1)^5\times(\mathbb{P}^2)^3$.}\label{CICYmat}
	\end{minipage}
    \begin{minipage}[b]{.45\textwidth}
	\centering
	\includegraphics[width=5cm]{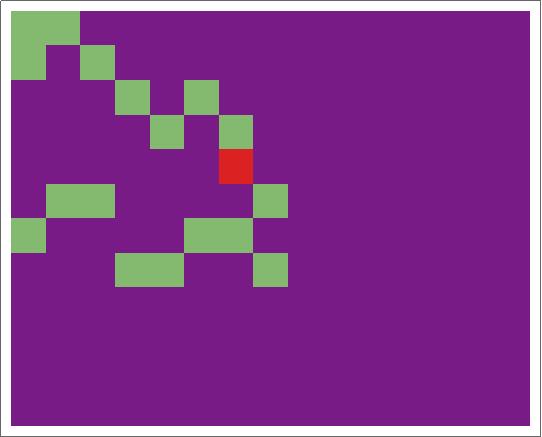}
	\caption{The corresponding image of Fig. \ref{CICYmat} where purple pixels are 0, green, 1 and red, 2.}\label{CICYex}
    \end{minipage}\hfill
\end{figure}

Let us now focus on the CICY dataset of 7890 inequivalent complete intersection CY 3-folds in products of (unweighted) complex projective spaces. As discussed in \S\ref{CICY3}, a CICY is represented by a matrix, whose entries are 0 to 5, with number of rows ranging from 1 to 12 and number of columns ranging from 1 to 15. In terms of computer graphics, it is a $12\times15$ pixelated image with 6 different colours (or 6 shades of grey in greyscale image). As an example, the CICY of 8 equations in $(\mathbb{P}^1)^5\times(\mathbb{P}^2)^3$ is the matrix in Fig. \ref{CICYmat} such that we have the image as in Fig. \ref{CICYex}.

As a matter of fact, we need CNN to take advantage of pixelations of CICYs. Nevertheless, for this task, we are only using the graphic images to emphasize that even though our computers have no knowledge of algebraic geometry, they can still ``learn" to make good predictions, and we will keep using MLP in our analysis.

\begin{figure}[h]
    \centering
    \includegraphics[width=12.5cm]{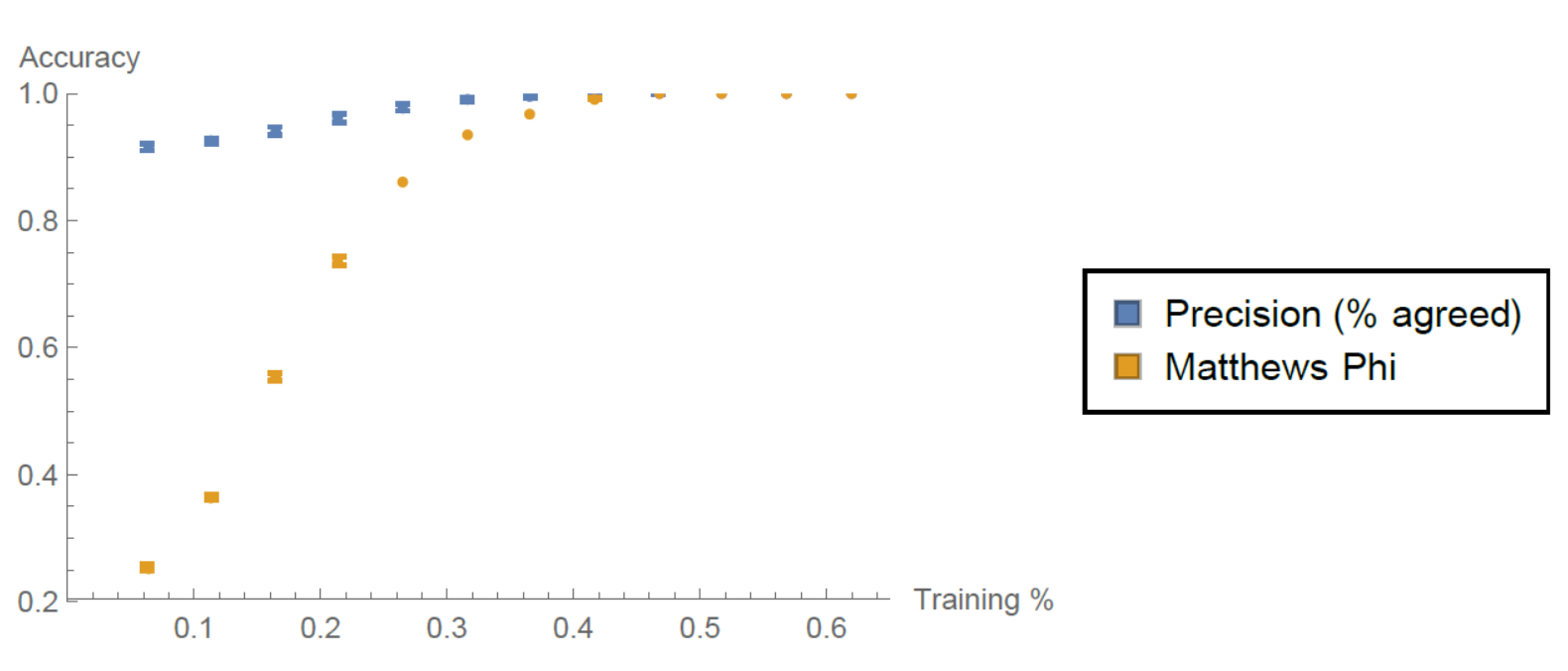}
    \caption{The learning curves for machine learning whether $h^{1,1}>5$ for CICYs.}
    \label{CICYcurve}
\end{figure}

Similar to \S\ref{warmupWP4}, the machine learns the binary query of the Hodge number $h^{1,1}>5$. The input would be $12\times15$ matrices, and the output is again either 0 or 1. Now we take 4000 random samples ($<50\%$) as our training data, and test the remaining 3890 data-points as validation. The learning time only takes about 5 minutes while the performance is remarkable. The accuracy $p$ is 97\% and the cosine distance $d_C$ reaches 0.98. For MCC, we have $\phi=0.87$. Varying the numbers of samples in $T$ yields the learning curves depicted in Fig. \ref{CICYcurve}.

We can see that there is a huge discrepancy between $p$ and $\phi$ for small $\gamma$'s. This is due to the great disparity between the sizes of two classes ($h^{1,1}\leq5$ and $h^{1,1}>5$). Indeed, Fig. \ref{fig:TopDatPlot} with the distribution of $h^{1,1}$'s verifies our argument. As aforementioned, MCC would be much more useful in this case. For larger $\gamma$'s, we do see the ascent of both curves, approaching to 1.

Let us now make our problem more sophisticated and compute the precise values of $h^{1,1}$. Here we try three different methods and compare their results:

\begin{itemize}
    \item NN Classifier: As $h^{1,1}\in[0,19]$, the output is a 20-channel classifier (cf. the 10-channel classifier in text recognition) with each neuron mapping to 0 or 1. The detailed architecture is decribed in \cite{CYLandscape, deeplearninglandscape, PLBpaper, bull}.
    \item NN Regressor: The output is some real number which means it is continuous. There are certain parameters known as \emph{hyperparameters} that need to be optimized before training by hand, such as the number of hidden layers etc. This is discussed in detail in Appendix C in \cite{bull}.
    \item SVM Classifier: The output is one of the possible values of $h^{1,1}$, that is, some integer between 0 and 19. The hyperparameter optimization is also discussed in \cite{bull}.
\end{itemize}

\begin{figure}[h]
    \centering
    \includegraphics[width=12.5cm]{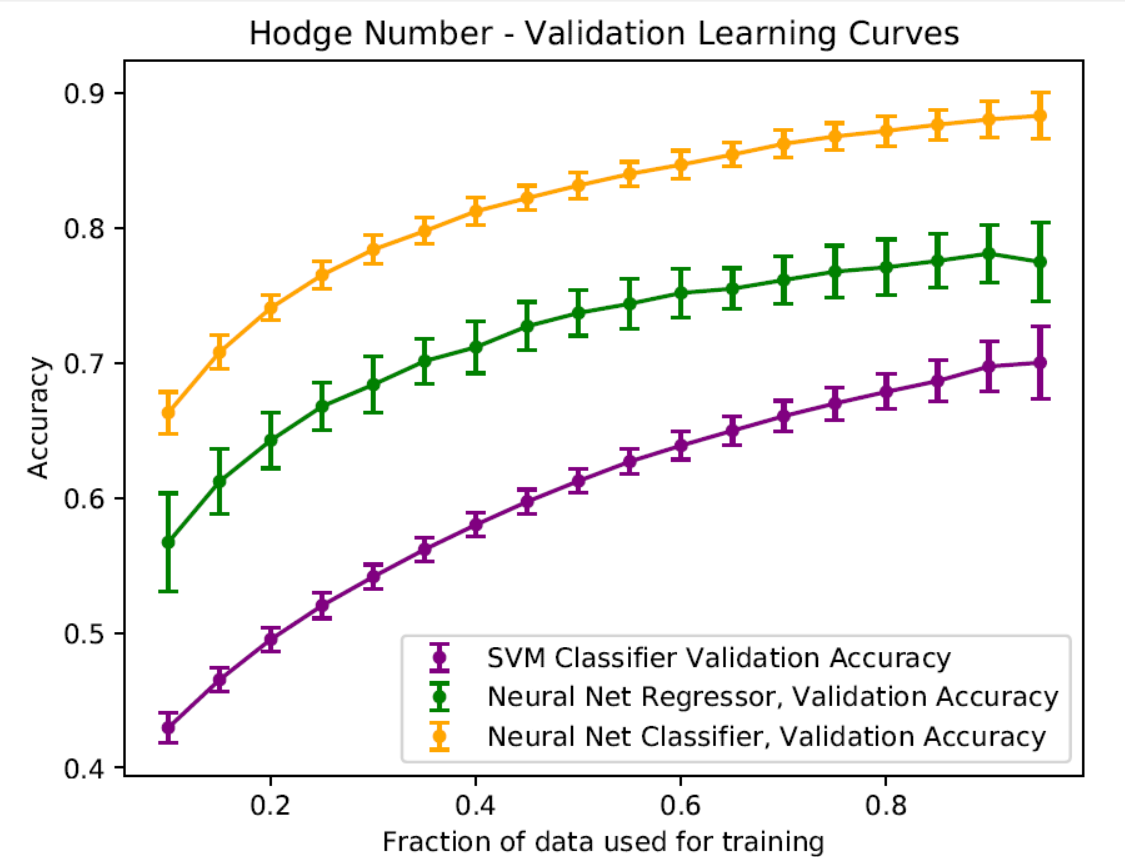}
    \caption{The learning curves generated by averaging over 100 different random cross validation splits.}
    \label{3curves}
\end{figure}

\begin{figure}[h]
    \centering
    \includegraphics[width=12.5cm]{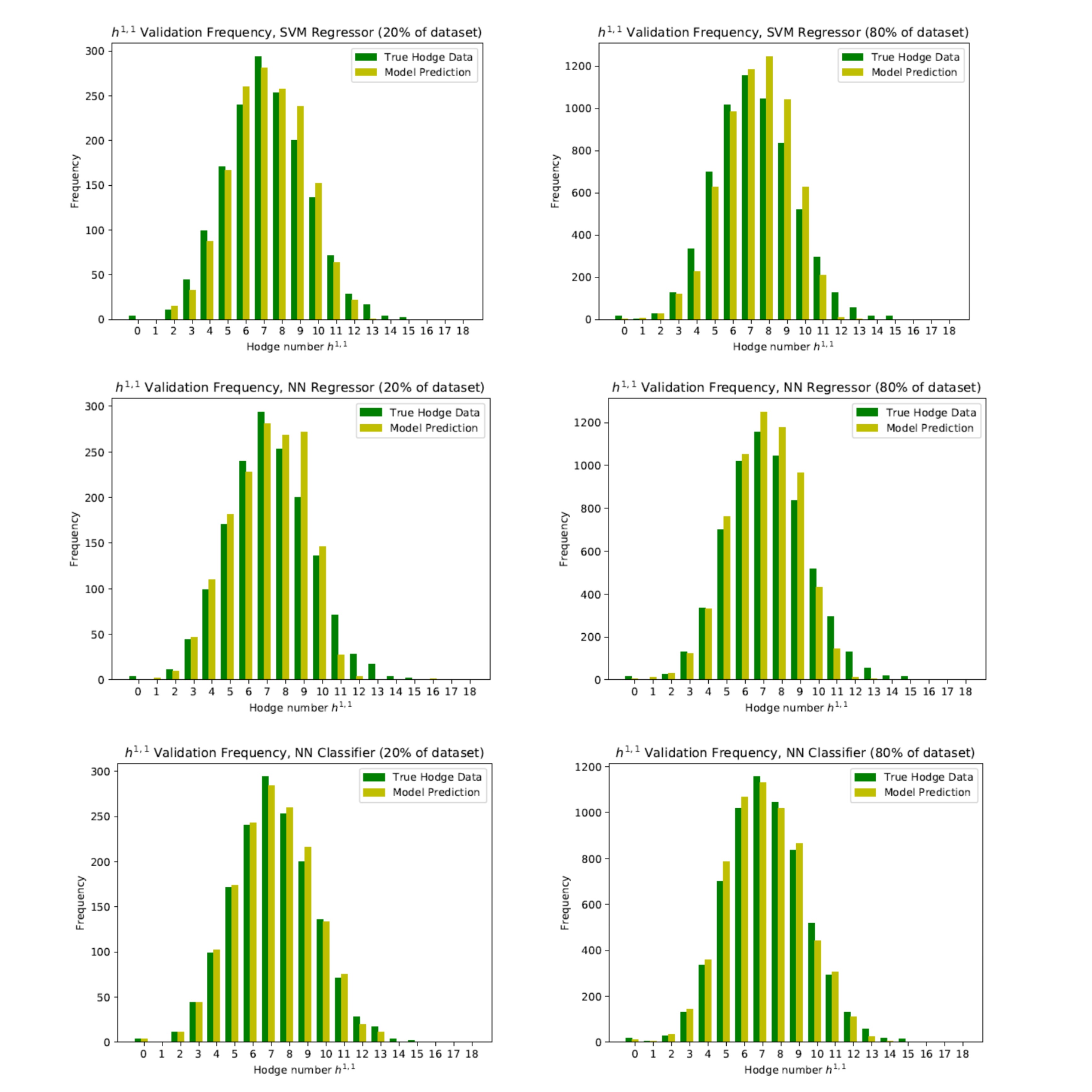}
    \caption{The frequencies of $h^{1,1}$'s.}
    \label{barchart}
\end{figure}

We can now plot the three learning curves as in Fig. \ref{3curves}.
We see that the NN classifier performs best in the machine-learning. Again, it is impressive that such training on an ordinary laptop takes only about 10 minutes and the validation only takes a few seconds. For reference, we plot the histograms of frequencies of predicted and actual $h^{1,1}$'s with validation sets of sizes 20\% and 80\% of the total data respectively for the three methods in Fig. \ref{barchart}.

\subsection{Distinguishing Elliptic Fibrations}
\label{ellipfib}

As CICYs may admit elliptic fibration (EF) structures, \cite{ellipticfibrations} machine-learnt these elliptic fibrations. Here we will contemplate the 643 CICY 3-folds with $h^{1,1}\leq4$ since all those with $h^{1,1}>4$ can be (obviously) elliptically fibred \cite{Gray:2014fla,Anderson:2016ler,Anderson:2016cdu,Anderson:2017aux}. As we have an unbalanced dataset (with 53 non-elliptic and 590 elliptic), we would like to make an enhancement on 53. Notice that CICY configurations are the same up to row and column permutations. We can therefore take 10 random permutations (independently) of both rows and columns on each of the 53 configuration matrices, which yields $10^2\times53=5300$ non-elliptic cases with output 0. We also perform 3 such permutations so that we have $3^2\times590=5310$ elliptic cases with output 1. Moreover, as these CICYs can all be represented by configuration matrices with 6 rows and 7 columns, the input will be a $6\times7$ matrix.

\begin{figure}[h]
    \centering
    \includegraphics[width=12.5cm]{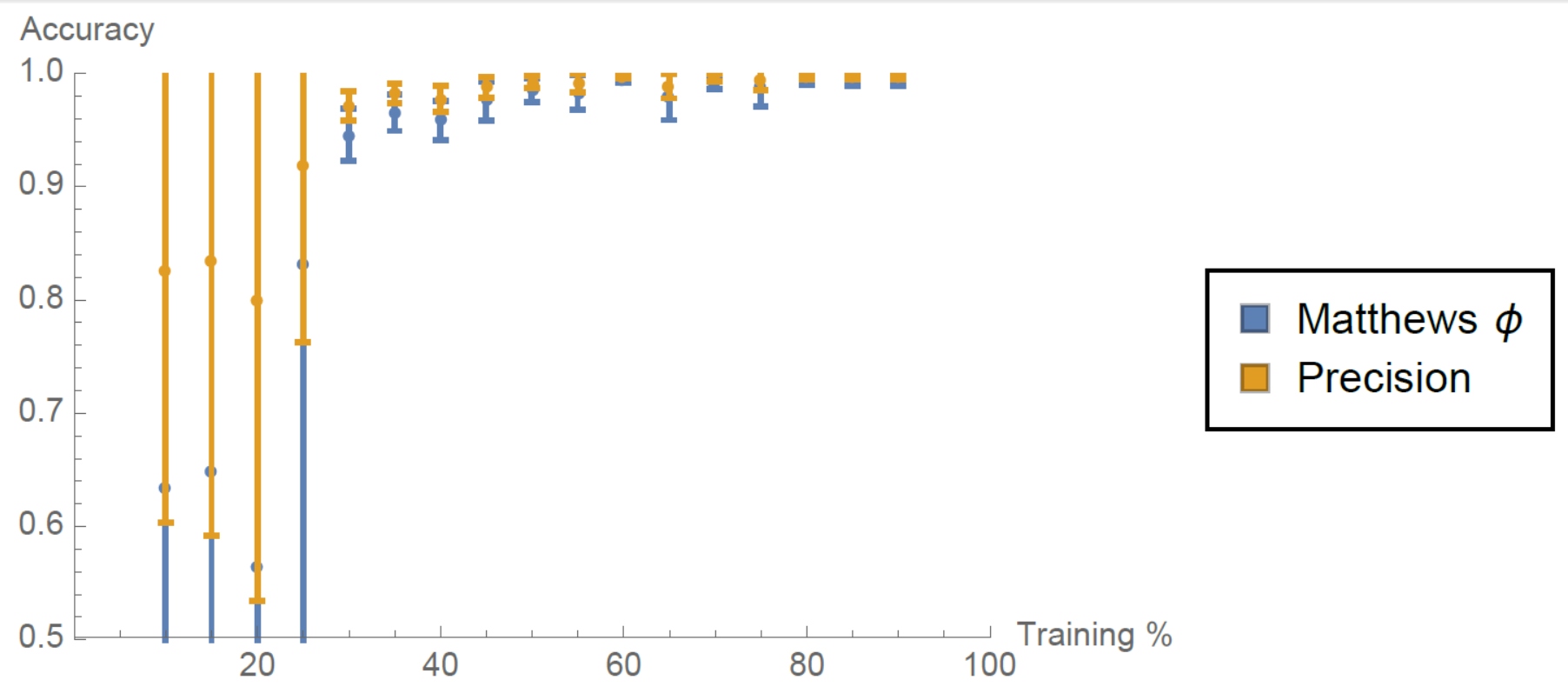}
    \caption{The learning curves for the (enhanced) ten thousand data-points on EFs.}
    \label{EF}
\end{figure}

Now we are dealing with our familiar binary queries. Following the similar recipe as above, we can draw the learning curves in Fig. \ref{EF}.
The error bars for up to about 25\% looks ugly as there is inadequate training data. However, from 30\% and above, machine-learning gives us a pretty nice result with high accuracy. Again, each training only takes a few seconds.

\begin{figure}[h]
    \centering
    \includegraphics[width=12.5cm]{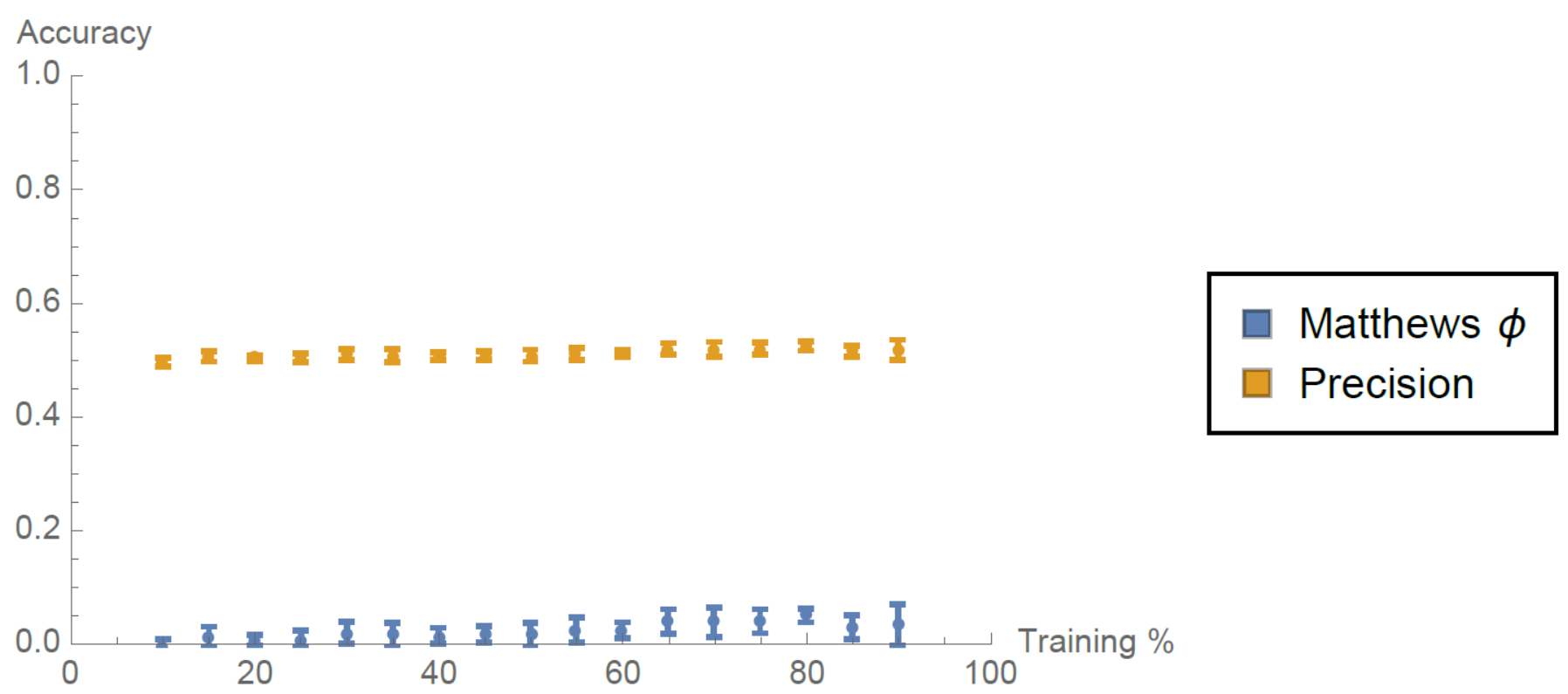}
    \caption{The learning curves on a control set of a randomly chosen property.}
    \label{control}
\end{figure}

It is also worth noting that we can make a \emph{control test} for this problem. We just arbitrarily choose 53 configuration matrices out of the 643 and assign 0 or 1 to these 53 matrices randomly. Then we have the learning curves depicted in Fig. \ref{control}.
We see that the machine-learning is poorly behaved (with $\sim50$\% precision and $\phi\sim0$) which shows that there is no inherent pattern for the machine to find in the control test. In contrast, EF is truly not a random property.

\section{A Digression: Group Theory}
\label{gptheory}

\subsection{Learning Cayley Tables}
\label{Cayleytab}

Now we would like to apply machine-learning to more basic problems in mathematics. Let us first start with recognizing Cayley tables $\mathcal{C}$ \cite{cayley} out of Latin squares $\mathcal{L}$.

\begin{figure}[h]
    \centering
    \includegraphics[width=12.5cm]{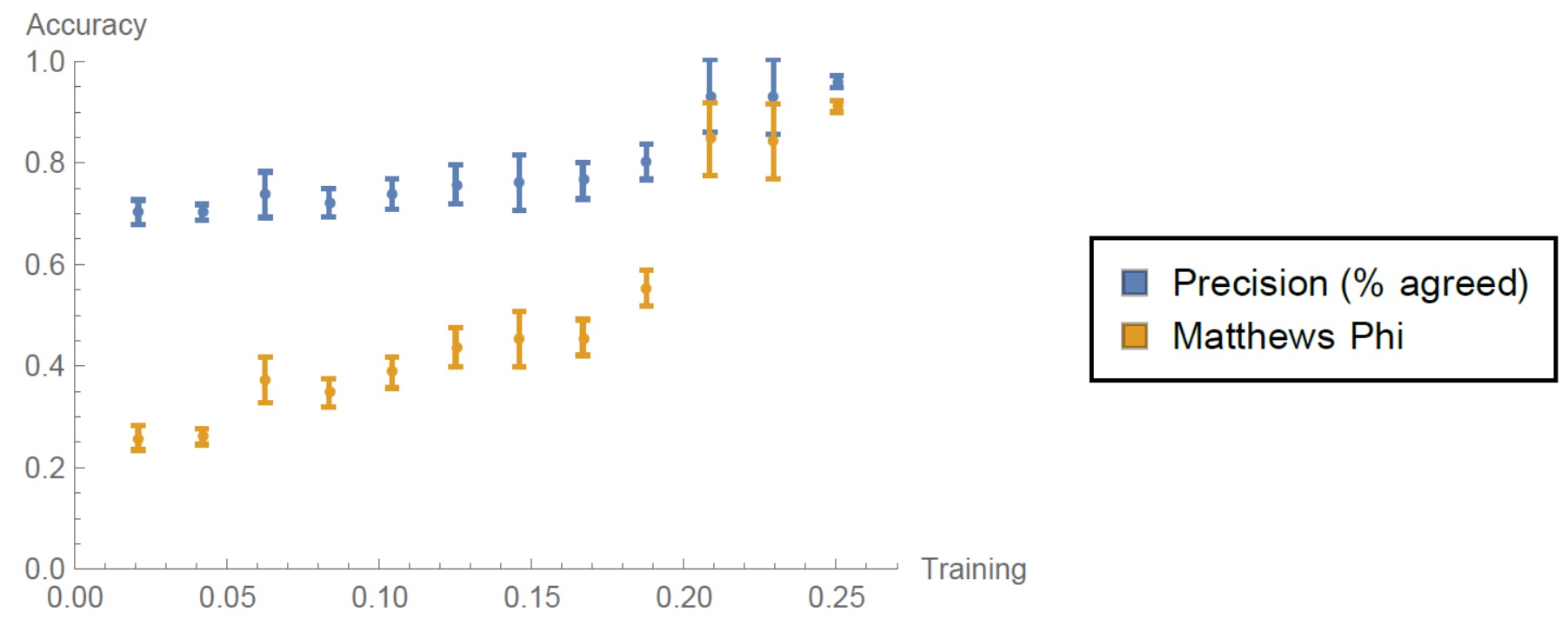}
    \caption{The learning curves for $n=8$ Latin squares. We vary the training size from 500 to 6000 in increments of 500.}
    \label{cayleycurve}
\end{figure}

Allowing permutations of rows and columns, the number of Cayley tables of size $n$ will be\footnote{If we naively consider all possible permutations, then $\#\mathcal{C}_n$ should be $(n!)^2\times\#G$ (where $\#G$ denotes the number of elements in group $G$.). However, Cayley tables always have degeneracies under permutations which leads to only $(n!)\times\#G$ distinct matrices though this is not obvious for non-symmetric matrices (which corresponds to non-abelian groups).} $\#\mathcal{C}_n=1,2,6,48,120,1440,\dots$ These Cayley tables form a subset of Latin squares\footnote{A Latin square is a $n\times n$ matrix filled by $n$ symbols (here $1,2,\dots,n$), each of which appears exactly once in each row and in each column.} $\mathcal{L}$. The number of Latin squares grows as:$\#\mathcal{L}_n=1, 2, 12, 576, 161280, 812851200, \dots$
Compare $\#\mathcal{C}$ with $\#\mathcal{L}$, we see that the probability of a Latin square being a Cayley table is \emph{essentially} 0 from $n$ as small as 5. This is important for our algorithm so that we can choose $n\geq5$, and assign $\mathcal{L}\rightarrow0$ and $\mathcal{C}\rightarrow1$. Thus, we are back to our familiar binary query. A more detailed descriptions of algorithms can be found in \cite{algstruc}. Here, we will just give out the learning curves as in Fig. \ref{cayleycurve}.
Both of the measures show that we have a perfect result when we have $\sim25\%$ out of the total data as training data.

\subsection{Learning Finite Simple Groups}
\label{simplegp}

After recognizing the Cayley tables, one may wonder how machine-learning would perform when studying other group properties. We now focus on the problem of finite simple groups.

\begin{figure}[h]
    \centering
    \includegraphics[width=12.5cm]{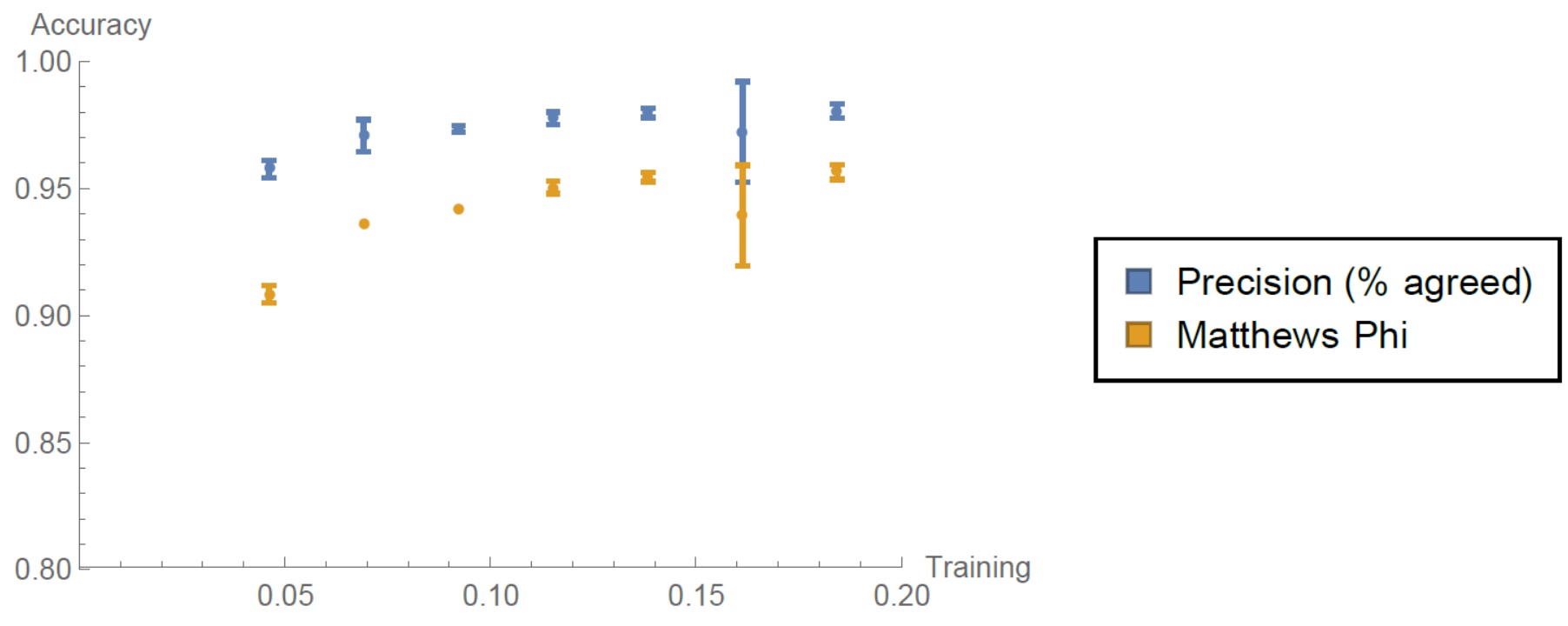}
    \caption{The learning curves for identifying whether a finite group of order $n\leq70$ is simple. There are 20 simple groups out of the total 602 groups.}
    \label{simpcurve}
\end{figure}

We know that it is not straightforward to determine whether a finite group is simple by just contemplating its Cayley table. However, there shouldn't be random properties in group theory, so we would like to see how machines can behave in this task. The detailed treatment can be found in \cite{algstruc}. Here, we report the learning curves as in Fig. \ref{simpcurve}.
We see that even at a low percentage of training data, the machine can still make very good predictions directly from Cayley tables without knowing Sylow theorems.

More examples of studying group properties and algebraic structures via machine-learning can be found in \cite{algstruc}. Recently, similar explorations have also been done in number theory \cite{Alessandretti:2019jbs}.

\section{Summary and Outlook}
\label{outlook}

As argued in \cite{deeplearninglandscape,PLBpaper}, any computational algebraic geometry problem is machine-learnable as it is in essence a finite number of steps finding kernels and cokernels of integer matrices. Thus, machines can be pretty well-behaved although they know nothing about algebraic geometry. On the other hand, this also shows that our properties in algebraic geometry are not random. Otherwise, our machine-learning would fail, just like the control case in \S\ref{ellipfib}.

Hence, we expect that machine so far is not able to learn number theory in which elusive prime numbers play a key role. As a sanity check, when we input a bunch of prime numbers and train our machine to valid larger primes in our data, we only achieve a terrible 0.1\% accuracy. After all, AI does not stand for all-powerful incredibility. Machine learning is good at matrix/tensor manipulations (and this is why $\mathtt{TensorFlow}$ is given such a name).

Anyway, algebraic geometry is the area where machine learning can show its power as we have seen above. Despite our success in machine-learning, we still do not know why this works. Unlike other disciplines in science, we do not know what is going on among neurons and their connections, and we can still get good predictions from them regardless of the theoretical intractability. An \emph{almost} 1 accuracy certainly cannot satisfy mathematicians, but machine learning still bypasses the expensive steps for practical purposes.

After decades of research, string landscape now solidly resides in the era of big data. CY manifolds is only a small portion of the heterotic landscape. There is plenty of room at the bottom where NNs can act as classifiers or predictors for generalized K\"ahler geometry, for stable holomorphic bundles, for quivers and brane tilings and so forth. The landscape is still on the threshold of benefiting from data science.

\section*{Acknowledgement}
We would like to thank Thomas Creutzig  and Steve Rayan for organizing the ``PIMS - USaskatchewan Summer School on Algebraic Geometry in High-Energy Physic'' which provided a wonderful atmosphere for mathematicians and physicists, students and experts, to interact and collaborate. YHH would like to thank STFC for grant ST/J00037X/1. EH would like to thank STFC for the PhD studentship.
	
\appendix

\section{Some Complex Geometry}
\label{complexgeometry}

Let us consider a complex manifold $\mathcal{M}$ of complex dimension $m$. Then the set of ($p$,$q$)-forms, $\Omega^{p,q}(\mathcal{M})$, obviously form an abelian group under addition. As $\mathcal{M}$ is also a $2m$-dimensional real manifold, we can decompose the familiar exterior derivative into two pieces: $\text{d}=\partial+\bar{\partial}$, such that $\partial$ acts on the holomorphic part of a ($p$,$q$)-form while $\bar{\partial}$ acts on the antiholomorphic part, that is, $\partial:\Omega^{p,q}(\mathcal{M})\rightarrow\Omega^{p+1,q}(\mathcal{M})$ and $\bar{\partial}:\Omega^{p,q}(\mathcal{M})\rightarrow\Omega^{p,q+1}(\mathcal{M})$. Followed from $\text{d}^2=0$, we get $\bar{\partial}^2=0$ (as well as $\partial^2=0$ and $\partial\bar{\partial}+\bar{\partial}\partial$=0). We may therefore construct the \emph{cochain complex}:
\begin{equation}
	0\stackrel{\bar{\partial}}{\longrightarrow}\Omega^{p,0}(\mathcal{M})\stackrel{\bar{\partial}}{\longrightarrow}\Omega^{p,1}(\mathcal{M})\stackrel{\bar{\partial}}{\longrightarrow}\dots\stackrel{\bar{\partial}}{\longrightarrow}\Omega^{p,m}(\mathcal{M})\stackrel{\bar{\partial}}{\longrightarrow}0.
\end{equation}
Then
\begin{definition}
	The \emph{Dolbeault cohomology group} is defined as
	\begin{equation}
		H^{p,q}_{\bar{\partial}}(\mathcal{M}):=\frac{\text{ker}(\bar{\partial}:\Omega^{p,q}(\mathcal{M})\rightarrow\Omega^{p,q+1}(\mathcal{M}))}{\text{Im}(\bar{\partial}:\Omega^{p,q-1}(\mathcal{M})\rightarrow\Omega^{p,q}(\mathcal{M}))}.
	\end{equation}
	The dimensions of the Dolbeault cohomology groups are known as the \emph{Hodge numbers}, $h^{p,q}:=\text{dim}H^{p,q}_{\bar{\partial}}(\mathcal{M})$.
\end{definition}
Not all the Hodge numbers are independent. For any complex manifold, Kodaira-Serre duality yields \cite{Griffiths}
\begin{equation}
	h^{p,q}=h^{m-p,m-q}.\label{serre}
\end{equation}
In particular, we have
\begin{equation}
	h^{0,0}=h^{m,m}=1.
\end{equation}
Also, one can prove that the Dolbeault cohomology group is always of finite dimension, viz, $h^{p,q}<\infty$.

\subsection{K\"ahler Manifolds}
\label{kahler}

As we will see, K\"ahler structure would put more constraints on the Hodge numbers. First, we need to introduce
\begin{definition}
	A \emph{Hermitian metric} $g$ on the complex manifold $\mathcal{M}$ with complex structure $J$ is a Riemannian metric satisfying $g(Jv,Jw)=g(v,w)$ for any vector fields on $\mathcal{M}$. Equivalently, in terms of (complex) components, $g_{\alpha\beta}=g_{\bar{\alpha}\bar{\beta}}=0$. Then the \emph{Hermitian form} is the 2-form defined by $\omega(v,w):=g(Jv,w)$. Equivalently, in terms of (complex) components\footnote{We use Latin indices for real coordinates and Greek ones for complex coordinates.}, $\omega_{ab}=ig_{\alpha\bar{\beta}}-ig_{\bar{\alpha}\beta}$. Therefore, $\omega$ is also a (1,1)-form.
\end{definition}
As the pure holomorphic and antiholomorphic components of the Hermitian metric vanish, it is not hard to see that $\omega_{ab}=-\omega_{ba}$.

Now we are able to define
\begin{definition}
	The Hermitian metric $g$ is \emph{K\"ahler} if d$\omega=0$, and $\omega$ is called a \emph{K\"ahler form}. A complex manifold is a \emph{K\"ahler manifold} if it admits a K\"ahler metric.
\end{definition}
In some literature, the K\"ahler manifold is defined as a complex manifold having a symplectic form (being bilinear, non-degenerate and antisymmetric). In fact, bilinearity and non-degeneracy come from the Riemannian metric, and antisymmetry follows the Hermicity of the metric as mentioned above.

It is worth remarking that since d$\omega=0$ (which is the same as $\partial_\alpha g_{\beta\bar{\gamma}}=\partial_\beta g_{\alpha\bar{\gamma}}$ along with its conjugate equation), then equivalently we can write $\omega=i\partial\bar{\partial}K$ for some real scalar function $K$ known as the \emph{K\"ahler potential}.

For K\"ahler manifolds, the Hodge numbers further satisfy
\begin{eqnarray}
	h^{p,q}&=&h^{q,p},\label{hodgesym}\\
	h^{p,p}&\geq&1.
\end{eqnarray}
Then (\ref{serre}) and (\ref{hodgesym}) yield
\begin{equation}
	h^{p,q}=h^{m-q,m-p}.
\end{equation}
If the manifold is further Calabi-Yau, one can show that $h^{m,0}=h^{0,m}=1$, and $h^{m,p}=h^{p,m}=h^{0,p}=h^{p,0}=0$ for $0<p<m$ \cite{hubsch_calabi-yau_1992}.

K\"ahler geometry is ubiquitous in physics. We care about K\"ahler manifolds since they preserve holomorphicity under parallel transportations of vectors, and a K\"ahler structure is essential for a manifold being Calabi-Yau.

\subsection{Chern Classes}
\label{chern}

Another important concept for Calabi-Yau manifolds is the \emph{Chern classes}.
\begin{definition}
	Given the complex vector bundle $E$ over the complex manifold $\mathcal{M}$ of complex dimension $m$ and the gauge group (aka structure group) $G$, let $F=\text{d}A+A\wedge A$ be the strength field (aka curvature 2-form) of the gauge potential (aka connection) $A$. We define the \emph{total} Chern class as\footnote{Although we use the 2-form $F$ in our definition, the Chern class should be independent of the choice of $F$ \cite{bouchard}.}
	\begin{equation}
		c(E)=\text{det}\left(I+\frac{i}{2\pi}F\right).
	\end{equation}
\end{definition}
Using $\text{det}\left(I+\frac{i}{2\pi}F\right)=\exp\left(\text{tr}\left(\log\left(I+\frac{i}{2\pi}F\right)\right)\right)$, we may expand the total Chern class as
\begin{equation}
	c(E)=c_0(E)+c_1(E)+c_2(E)+\dots+c_m(E),
\end{equation}
where $c_k(E)$ is the $k^{\text{th}}$ Chern class\footnote{Strictly speaking, they are \emph{Chern forms}, and the Chern classes are the cohomology classes of the Chern forms. When $E$ is the holomorphic line bundle $T^{1,0}\mathcal{M}$, we also say that $c_k$(E) is the Chern class of the manifold $\mathcal{M}$ and denote it as $c_k(\mathcal{M})$.}. As $F$ is a 2-form, $c_k(E)$ is a $2k$-form and vanishes for $k>m$. There are some explicit formulae for the Chern forms such as
\begin{eqnarray}
	&&c_0(E)=1,\\
	&&c_1(E)=\frac{i}{2\pi}\text{tr}(F),\\
	&&\dots\nonumber\\
	&&c_m(E)=\left(\frac{i}{2\pi}\right)^m\text{det}(F).
\end{eqnarray}
For more on Chern classes (including Chern characters, Todd classes etc.), see \cite{Nakahara}.

\section{Toric Varieties}\label{toric_varieties}
	
	Toric varieties are a generalisation of complex weighted projective vector spaces. Complex projective space, $\mathbb{CP}^m$, is the $\mathbb{C}^{m+1}$ complex space with the origin removed, and quotiented out by the identification:
	\begin{equation}
	(z_0,z_1,...,z_m) \sim (\lambda z_0,\lambda z_1,...,\lambda z_m) \quad \forall \lambda \in  \mathbb{C}\setminus\{0\}\,;
	\end{equation}
	such that all points along lines through the origin are identified. This concept is generalised to a weighted complex projective space, denoted $\mathbb{CP}^{(a_0,a_1,...,a_m)}$, where instead the identification quotiented with is:
	\begin{equation}
	(z_0,z_1,...,z_m) \sim (\lambda^{a_0} z_0,\lambda^{a_1} z_1,...,\lambda^{a_m} z_m) \quad \forall \lambda \in \mathbb{C}\setminus\{0\}\,,\label{weightedproj}
	\end{equation}
	with the $a_i$ constants acting as powers on $\lambda$. Both weighted and unweighted complex projective spaces are types of toric variety, however more toric varieties can be formed through more general removal of a subset of the space, $\mathcal{U}$, and quotienting by a more general algebraic torus $(\mathbb{C}\setminus\{0\})^p$. Thereby a toric variety is defined: 
	\begin{equation}
	\mathcal{M} = (\mathbb{C}^m \, \setminus \, \mathcal{U}) \,/\, (\mathbb{C}\setminus\{0\})^p\,;
	\end{equation} 
	where there are $p$ identifications to quotient out by with $p$ coefficient sets of non-zero complex numbers. Note for unweighted complex planes all coefficients in the set are the same \cite{Skarke_toric}.
	
	Toric varieties can also be defined in terms of fans. Whereby a fan is a collection of cones such that all cone faces are themselves cones in the fan, and cones intersect at mutual faces. A toric variety is then defined using the one-dimensional cones in a fan defined on a vector space obtained from a lattice ($\mathbb{Z}^n\otimes_\mathbb{Z}\mathbb{R}$). First a complex homogeneous coordinate is associated to each generator of a one-dimensional cone such that a fan with $k$ generators, for the $k$ one-dimensional cones, corresponds to $\mathbb{C}^k$. Next, all points in the space which correspond to combinations of these generators which are not contained within cones in the fan are removed from the space. Finally the remaining space is quotiented by equivalence relations that correspond to these generator combinations outside of the fan. More mathematically this defines the full toric variety as:
	\begin{equation}
	\mathcal{M} = \{\mathbb{C}^k \, \setminus \mathcal{U}\}\, /\, \mathcal{G}\,;
	\end{equation}
	for $\mathcal{U}$ as the set of generator combinations that are not contained within the fan; and $\mathcal{G}$ the algebraic torus quotiented by (potentially with some additional finite abelian group) to give the equivalence relations, usually $(\mathbb{C}\setminus \{0\})^p$.
	
	This construction method can be more practical, as singularities in a manifold correspond to singularities in its corresponding toric variety; and singularities in a toric variety can be resolved by introducing further cones into the variety's fan in a specific way.
	
	A general toric variety is Calabi-Yau if its generators exist in an affine hyperplane of the lattice. This makes identifying Calabi-Yau manifolds very easy from the fan structure. The Calabi-Yau property can alternatively be identified from the charges of the equivalences quotiented out by in the variety definition. If the sum of the coefficients (``charges'') for each equivalence relation is zero for all the relations, then the space is also Calabi-Yau. Since a Toric variety is compact if its fan fills the lattice space; by definition all Calabi-Yau manifolds formed from toric varieties are thus non-compact, as their generators do not span the lattice (but exist in a codimension 1 hyperplane of it).
	
	Toric diagrams, which are useful for manifold classification and physical interpretation, can then be constructed. Considering the hyperplane containing the Calabi-Yau generators in the lattice space, a graph can be formed by connecting the points where this hyperplane intersections with the fan's generators. The dual of this graph is defined to be the manifold's ``toric diagram'', and it encodes the degeneration of the fibres of the manifold.
	
\begin{figure}[h!]
		\centering
		\includegraphics[width=75mm]{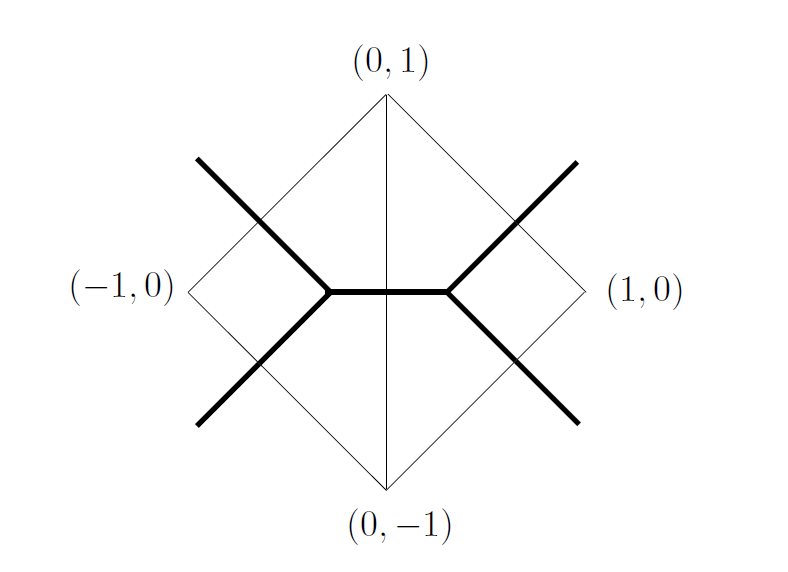} 
		\caption{An example of a toric diagram (in bold) for the resolved conifold \cite{bouchard}. The points give the intersection of the one-dimensional cone generators of the fan with the hyperplane they all exist in under the Calabi-Yau condition. Connecting the points gives the graph, whose dual is the toric diagram.}
		\label{toric_dia}
	\end{figure}

	The toric diagram in figure \ref{toric_dia} represents a new branch of Calabi-Yau surfaces, known as conifolds. These permit conic singularities in their description, and were key in deriving mirror symmetry which connects Calabi-Yau 3-folds. The charges defining this conifold are $Q_i = (1,1,-1,-1)$, since these sum to zero the space is thus Calabi-Yau. Generators are created that all exist in the same hyperplane (of the form $(v_i,1)$ for $v_i$ a 2-dimensional vector), and satisfy the relation $\sum_i Q_i \cdot v_i = 0$. This gives the $v_i$'s as the points labelled in figure \ref{toric_dia}, where they are drawn as points in the $x_3=1$ hyperplane. 
	
	The full $(v_i,1)$ vectors define the 1-dimensional cone generators of the fan in the lattice space. In the corresponding complex space, where each of these generators has a coordinate associated to it, the above conifold example is thus defined by the relation:
	\begin{equation}
	\sum_i Q_i \cdot |z_i|^2 = |z_1|^2 + |z_2|^2 - |z_3|^2 - |z_4|^2 = t\,,
	\end{equation}
	for a parameter $t$ known as the K\"ahler parameter (which counts the number of K\"ahler forms on the manifold). Considering the conifold as a fibration of a base space with a $T^3$ torus, this relation along with the other boundary relations $|z_i|^2 = 0$ define the base space of the conifold total space. The intersections of the hyperplanes defined by these equations gives an equivalent version of the toric diagram, and the full conifold space is the base with its fibration. This is where the conic idea behind the ``conifold'' name comes from \cite{bouchard}.

\section{Introduction to Machine Learning}
\label{introduction}

\subsection{Text recognition}
\label{textreg}

To make a start, let us first contemplate a prototypical example of text recognition \cite{CYLandscape, deeplearninglandscape, PLBpaper}. Given the 10 handwritten digits:
\begin{equation}
\includegraphics[width=10cm]{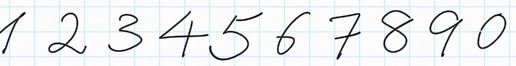},
\end{equation}
we would want the computers to recognize them. Notice that the inputs, which are images, are essentially $m\times n$ matrices representing the pixels in our 2-dimensional grid. The entries are either 0 or 1, encoding the black-white information for our binary images\footnote{If we want colour images, then the entries of these matrices would range from 0 to 1, indicating the percentage of RGB values.}. The outputs are simply the ten integers from 0 to 9 (aka \emph{10-channel outputs}).

For physicists and mathematicians, it is natural to think of the hardcore approach to solve this problem by finding the Morse function via detecting the different critical points for different digits. We are able to do so because the shapes vary from one digit to another. However, there will be two cons: the variation of different hand-writings and the too expensive computation.

This is pretty much the situation we have in algebraic geometry. For instance, the Gr\"obner basis is way too expensive for computation, and the input may also vary in configuration \cite{CYLandscape,deeplearninglandscape,PLBpaper}.

Computer scientists and data scientists tell us that we can machine-learn this problem as the following steps:
\begin{enumerate}
	\item Data Aquisition: We collect adequate known cases (input$\rightarrow$output), which are called \emph{training data}. For instance, the National Institute of Standards and Technology (NIST) database \cite{NIST} has $\sim10^6$ samples in the form:
	\begin{equation}
	\includegraphics[width=11cm]{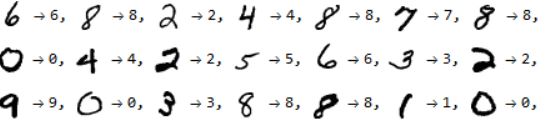}\dots
	\end{equation}
	
	\item Neural Network (NN) Setup and Training: This is the core machine-learning part that we will fixate on in \S\ref{nn}. If the NN is sufficiently complex, we will call it deep learning.
	
	\item Validation: After our machine/AI has ``learnt" the training data, we can feed it with unseen data named \emph{validation data} in the same form as training data. This test will then reflect how the machine performs after training.
	
	\item Prediction: If the NN passes our validation test, then it can be used in applications.
\end{enumerate}

Similar to the example above, a typical problem in string theory and computational algebraic geometry has the format:
\begin{equation}
\begin{tikzpicture}
[rec/.style={rectangle, draw=black, fill=white, minimum width=3cm, minimum height=1cm},]
\node at (0,0){INPUT};
\node[rec] (a) at (0,-1){Integer Tensor};
\node at (4,0){OUTPUT};
\node[rec] (b) at (4,-1){Integer};
\draw[->] (a.east)--(b.west);
\end{tikzpicture}.
\end{equation}
As a FLOSS (Free/Libr\'e and Open Source Software) with pseudo-code nature, $\mathtt{Python}$ is simple and popular. Hence, $\mathtt{Python}$ becomes our preference\footnote{Beginners can refer to \cite{bytepython} which is a well-known free book on $\mathtt{Python}$.}. There are also standard softwares for our $\mathtt{Python}$ programming language such as $\mathtt{SageMath}$ (aka $\mathtt{Sage}$ or $\mathtt{SAGE}$) \cite{sagemath} and $\mathtt{TensorFlow}$ \cite{tensorflow2015-whitepaper}. Perhaps, the only shortage of $\mathtt{Python}$ is that it is not as fast as $\mathtt{C}$ or $\mathtt{Java}$. Fortunately, we are allowed to run $\mathtt{Python}$ on the $\mathtt{Java}$ platform thanks to $\mathtt{Jpython}$/$\mathtt{Jython}$\footnote{We also have $\mathtt{Cython}$ as a compiler which compiles our $\mathtt{Python}$ codes to $\mathtt{C}$. Notice, however, $\mathtt{Jython}$ and $\mathtt{Cython}$ are different in principle. In particular, speeding up is pretty much the goal for the latter while it is not for $\mathtt{Jython}$ which mainly aims to import and use $\mathtt{Java}$ classes. Anyway, both of them make our codes run faster and get rid of the heavy codes in the meantime.}.

\subsection{Neural Networks}
\label{nn}

Motivated by biological neural networks, \emph{(artificial) neural networks ((A)NNs)} are essentially a set of algorithms designed to find patterns/relationships of our input data. They cluster and label the data we feed them. In the end, they return numerical results interpretting the patterns they recognize.

The basic components of NN are units/nodes called \emph{neurons} analogous to human brains. As we will see below, a neuron in the network is a \emph{function} that collects and classifies data. The neurons are often organized in \emph{layers}, and there may be multiple connections among different layers. A large collection of neurons then gives rise to an NN. The more layers or the greater the complexity of the inter-connectivity we have, the deeper our learning is.
\begin{definition}
	A single-layer \emph{perceptron} (SLP) is an archetypal neuron which is a function called \emph{activation function}, $f(x_i)$, of some input vector $x_i$. The activation function is typically taken to be binary (i.e. Heaviside function), or a sigmoid function such as the hyperbolic tangent or the logistic function\footnote{We choose sigmoid functions because they lie between 0 to 1, hence useful in probability predictions. Besides, they are differentiable with relatively simple derivatives proportional to themselves.}. The activation function is set to contain real parameters of the form $f(\sum\limits_iw_ix_i+b)$, where $w_i$'s and $b$ are called \emph{weights} and \emph{bias} respectively.
\end{definition}

Another widely used function is the famous rectified linear unit (ReLU) activation function:
\begin{equation}
f(x):=\left\{
\begin{array}{rcl}
x,  &\ &x>0; \\
0,  &\ &x\leq0.  
\end{array}
\right.
\end{equation}
As a result, this is less computationally expensive than sigmoid functions. The sparity of ReLU makes it behave more like a biological brain. Moreover, it has a better gradient propagation compared to sigmoid functions who has tiny gradients towards the ends. If the gradient is too small or even vanished, then the NN would learn rather slowly or even refuse to learn further. Although the gradient of ReLU will vanish for negative $x$, no activation function is always perfect for every case, and we need to choose the one fits best.

Now given some training data
\begin{equation}
T=\{(x^{(j)}_i,d^{(j)})\},
\end{equation}
where $x^{(j)}_i$'s are inputs and $d^{j}$'s are \emph{known} outputs with labelling $j$. Then we want our error, the standard deviation
\begin{equation}
SD=\sum_j\left(f\left(\sum_iw_ix_i^{(j)}+b\right)-d_i\right)^2,
\end{equation}
to be minimized with respect to $w_i$'s and $b$ by the method of steepest descent\footnote{This is often done numerically due to the large number of parameters.}. After such training, we can now proceed to validation against unseen data. As we can see, this is basically the (non-linear) regression for model function $f$.

As aforementioned, we can have many layers for deep learning. This leads to
\begin{definition}
	A \emph{multi-layer perceptron} (MLP) is a sequence of layers where the output of the previous layer is the input of the next layer, with different weights and biases. The output of the $i^{\text{th}}$ neuron in the $n^{\text{th}}$ layer is
	\begin{equation}
	f_i^n=f(W_{ij}^nf_j^{n-1}+b_i^n).
	\end{equation}
	Notice that we have promoted the weights $w_j$'s to a (layer-wise) weight matrix $W^n$ such that the $W_{ij}^n$ denotes the weight $w_j$ connected from neuron $j$ in the $(n-1)^{\text{th}}$ layer to neuron $i$ in the $n^{\text{th}}$ layer. Likewise, the bias $b$ has become a bias vector $b_i$. Our input would be $f^0_i=x_i$.
\end{definition}
Thus, the MLP is depicted as
\begin{equation}
\begin{tikzpicture}
[round/.style={circle, draw=black, fill=white, minimum size=6mm},]
\node[round] (a1) at (0,0){};
\node[round] (a2) at (0,-1){};
\node[round] (a3) at (0,-2){};
\node at (0,-2.5){$\vdots$};
\node[round] (b1) at (2,-0.5){};
\node[round] (b2) at (2,-1.5){};
\node at (2,-2.5){$\vdots$};
\node[round] (c1) at (4,-0.5){};
\node[round] (c2) at (4,-1.5){};
\node at (4,-2.5){$\vdots$};
\node[round] (d1) at (6,-1){};
\node at (4.8,-1){$\dots$};
\node at (5.5,-1){$\vdots$};
\draw[-] (a1.east)--(b1.north west)
(a1.south east)--(b2.north west)
(a2.north east)--(b1.west)
(a2.south east)--(b2.west)
(a3.north east)--(b1.south west)
(a3.east)--(b2.south west)
(b1.east)--(c1.west)
(b1.south east)--(c2.north west)
(b2.north east)--(c1.south west)
(b2.east)--(c2.west)
(5.4,-0.5)--(d1.north west)
(5.4,-1.5)--(d1.south west);
\node at (1,-2.5){$W_{ij}^1,b_i^1$};
\node at (3,-2.5){$W_{ij}^2,b_i^2$};
\end{tikzpicture},
\end{equation}
where the left-most layer is the input layer while the right-most layer is the output layer. The layers between them are called hidden layers. This is just the simplest NN which only involves forward propagation from left to right. NNs can allow backward propagation and cycles as well. Besides MLP, we also have other NNs such as convolutional neural network (CNN) \cite{CYLandscape, deeplearninglandscape, PLBpaper, Goodfellow-et-al-2016}. Based on disscusions before, we have
\begin{definition}
	A \emph{convolutional (neural) network} is a neural network with general matrix multiplications replaced by convolutions in at least one of its layers.
\end{definition}
For a simplest case with no hidden layers, the output (aka \emph{feature map}) $s$ is
\begin{equation}
s=(x*w)(T)=\int x(t)w(T-t)dt,
\end{equation}
where $x$ is the input and $w$ is the weighting function called \emph{kernel} that restricts our input. For instance, we would like to learn how a stretch of coastline varies over a period of time. Then our input would be the position of the coastline with variable $T$ being the time we make each measurement. Moreover, measurements at later times (indicated by $t$) are more relevant, hence acquire more weights controlled by the kernel.

This is basically how CNNs use filters when scanning images. Instead of detecting images pixel by pixel, CNNs, just like what humans do, ``look at" an area known as \emph{receptive field} of an image. The movement of filter is measured by \emph{stride}. If the stride is 2, then the filter will move 2 pixels each time. Hence, CNNs are very powerful dealing with grahic inputs. Thus, for a 2-dimensional image as our input $x$, we want to use convolutions over both of the two axes. Then
\begin{equation}
S(i,j)=(x*w)(i,j)=\sum_m\sum_nx(m,n)w(i-m,j-n),
\end{equation}
where the kernel $w$ is now of dimension 2. As the dummy variables often varies less, we can \emph{flip} the kernel due to commutativity of convolutions:
\begin{equation}
S(i,j)=(w*x)(i,j)=\sum_m\sum_nw(m,n)x(i-m,j-n).
\end{equation}
Quite often, people do not bother flipping the kernel and use the \emph{cross-correlation}\footnote{It is worth noting that some literature call cross-correlations convolutions as well.} instead:
\begin{equation}
S(i,j)=(w*x)(i,j)=\sum_m\sum_nx(i+m,j+n)w(m,n).
\end{equation}

Besides the convolutional layers (and ReLU layers), a CNN also has \emph{pooling} layers that combine the outputs of neurons in one layer into a single neuron in the next layer and fully connected layers to connect  every neuron in one layer to every neuron in the next layer.

For more detailed stuff on NN, there are good texts such as \cite{NN1, NN2, NN3}.

\subsection{Support Vector Machines}
\label{svm}

Besides NN approach to machine learning, we also have support vector machines (SVMs), decision trees, $k$-nearest neighbours ($k$-NNs) etc. Here we are going to introduce the most widely used SVMs. SVMs, which take a more geometric approach compared to NNs, can act as both classifiers and regressors\footnote{In brief, programmes are asked to specify which categories the inputs belong to in classifications, and programmes need to predict a numerical value given some input in regressions. Hence, for classifiers, the function maps from input variables $x_i$ to \emph{discrete} output variables $y_i$ while for regressors, the function maps from input variables $x_i$ to \emph{continuous} output variables $y_i$. Besides classifications and regressions, there are other tasks in machine learning as well. See \cite{Goodfellow-et-al-2016}.}.

When an SVM acts as a classifier, it establishes an optimal hyperplane in the $n$-dimensional \emph{feature space} where the input $n$-vectors live in, indicating the binary feature as required in a classification predictive modelling. We can define the hyperplane as
\begin{equation}
\{\bm{x}\in\mathbb{R}^n|f(\bm{x})=\bm{w}\cdot\bm{x}+b=0\},
\end{equation}
where $\bm{w}$ is the normal vector to the hyperplane. Then
\begin{definition}
	The \emph{support vectors} are the points in the feature space lying closest to the hyperplane on either side denoted as $\bm{x}_i^{\pm}$. The \emph{margin} $M$ is the distance between these two vectors projected along $\bm{w}$, viz,
	\begin{equation}
	M:=\bm{w}\cdot(\bm{x}_i^+-\bm{x}_i^-)/|\bm{w}|.
	\end{equation}
\end{definition}
Our optimal hyperplane will maximize the margin. The reason is that we want our points in different classes as far from the separating hyperplane as possible since points close to the boundary can be easily misclassified. As the hyperplane is invariant under rescaling ($f=0\rightarrow\alpha f=0$), we can rescale $\bm{w}$ such that $f(\bm{x}_i^{\pm})=\pm1$ and $M=2/|\bm{w}|$. Thus, maximizing the margin is realized via minimizing $|\bm{w}|$. Moreover, as a classifier, the SVM gives $y_i$=1 if $\bm{w}\cdot\bm{x}_i+b\geq1$ while $y_i$=-1 if $\bm{w}\cdot\bm{x}_i+b\leq-1$. Equivalently, this yields $y_i(\bm{w}\cdot\bm{x}_i+b)\geq1$ for each $i$. Therefore, our problem is actually to minimize $|\bm{w}|$ with the constraints $y_i(\bm{w}\cdot\bm{x}_i+b)\geq1$. This can be solved using Lagrange multipliers\footnote{We use $|\bm{w}|^2$ rather than $|\bm{w}|$ here so as to get rid of the square root in the norm to make our life easier.}:
\begin{equation}
L=\frac{1}{2}|\bm{w}|^2-\sum_i\alpha_i(y_i(\bm{w}\cdot\bm{x}_i+b)-1).
\end{equation}
Hence,
\begin{eqnarray}
\frac{\partial L}{\partial\bm{w}}&=&\bm{w}-\sum_i\alpha_iy_i\bm{x}_i=0;\nonumber\\
\frac{\partial L}{\partial b}&=&-\sum_i\alpha_iy_i=0.
\end{eqnarray}

\begin{figure}[h]
	\centering
	\tikzset{every picture/.style={line width=0.75pt}} 
	\begin{tikzpicture}[x=0.75pt,y=0.75pt,yscale=-1,xscale=1]
	\draw  (112,260.16) -- (481.3,260.16)(148.93,24) -- (148.93,286.4) (474.3,255.16) -- (481.3,260.16) -- (474.3,265.16) (143.93,31) -- (148.93,24) -- (153.93,31)  ;
	\draw    (425.3,105.4) -- (200,217) ;
	\draw  [dash pattern={on 4.5pt off 4.5pt}]  (434.3,124.4) -- (209,236) ;
	\draw  [dash pattern={on 4.5pt off 4.5pt}]  (412.3,89.4) -- (187,201) ;
	\draw    (442.3,86.4) -- (442.3,124.6) ;
	\draw    (442.3,86.4) -- (449.3,86.6) ;
	\draw    (435.3,86.2) -- (442.3,86.4) ;
	\draw    (442.3,105.5) -- (449.3,105.7) ;
	\draw    (435.3,105.3) -- (442.3,105.5) ;
	\draw    (442.3,124.6) -- (449.3,124.8) ;
	\draw    (435.3,124.4) -- (442.3,124.6) ;
	\draw (240,180) node  [align=left] {{\LARGE \textcolor[rgb]{0.07,0.05,0.94}{.}}};
	\draw (260,195) node  [align=left] {{\LARGE \textcolor[rgb]{0.07,0.05,0.94}{.}}};
	\draw (220,195) node  [align=left] {{\LARGE \textcolor[rgb]{0.07,0.05,0.94}{.}}};
	\draw (312.65,161.2) node  [align=left] {{\LARGE \textcolor[rgb]{0.07,0.05,0.94}{.}}};
	\draw (340,151) node  [align=left] {{\LARGE \textcolor[rgb]{0.07,0.05,0.94}{.}}};
	\draw (275,164) node  [align=left] {{\LARGE \textcolor[rgb]{0.07,0.05,0.94}{.}}};
	\draw (381,110) node  [align=left] {{\LARGE \textcolor[rgb]{0.07,0.05,0.94}{.}}};
	\draw (407,109) node  [align=left] {{\LARGE \textcolor[rgb]{0.07,0.05,0.94}{.}}};
	\draw (351,96) node  [align=left] {{\LARGE \textcolor[rgb]{0.07,0.05,0.94}{.}}};
	\draw (381,170) node  [align=left] {{\LARGE \textcolor[rgb]{0.07,0.05,0.94}{.}}};
	\draw (456,85) node  [align=left] {$\epsilon$};
	\draw (460,124) node  [align=left] {$-\epsilon$};
	\draw (458,106) node  [align=left] {0};
	\draw (346,107) node  [align=left] {\{};
	\draw (386,162) node [xslant=0.02] [align=left] {\}};
	\draw (335,108) node  [align=left] {$\xi$};
	\draw (397,162) node  [align=left] {$\xi^*$};
	\end{tikzpicture}
	\caption{The ``soft margin" setting for linear SVR.}
	\label{softsvr}
\end{figure}
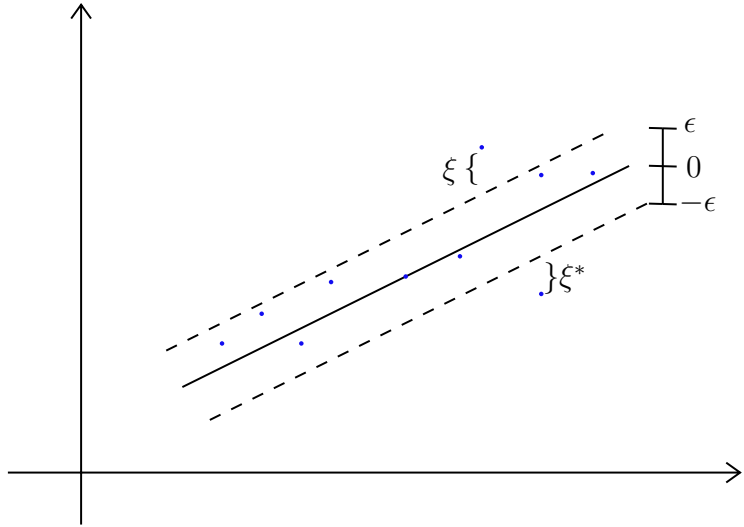

Likewise, the (linear) SVM regressor (aka support vector regression, SVM) follows the same principles, only with the difference of the output being real which make it difficult to make predictions. Now we have a tolerance of errors that allows $f(\bm{x})=\bm{w}\cdot\bm{x}+b$ to deviate from the actual result $y_i$ at most $\epsilon$. Meanwhile, we want to keep $f$ as flat as possible. As $|\nabla f|^2=|\bm{w}|^2$, our problem boils down to minimize $|\bm{w}|^2/2$ subject to the condition $|y_i-(\bm{w}\cdot\bm{x}_i+b)|\leq\epsilon$. This can be solved using Lagrange multipliers as well:
\begin{equation}
L=\frac{1}{2}|\bm{w}|^2-\sum_i\alpha_i(y_i-(\bm{w}\cdot\bm{x}_i+b)+\epsilon)+\sum_i\alpha^*_i(y_i-(\bm{w}\cdot\bm{x}_i+b)-\epsilon).\label{lagrangian}
\end{equation}
Hence,
\begin{eqnarray}
\frac{\partial L}{\partial\bm{w}}&=&\bm{w}-\sum_i(\alpha_i-\alpha^*_i)\bm{x}_i=0;\nonumber\\
\frac{\partial L}{\partial b}&=&-\sum_i(\alpha_i-\alpha^*_i)=0.\label{lagranderiv}
\end{eqnarray}
Notice that we actually require that such $f$ does exist with precision $\epsilon$ for all data-points. Sometimes, this may not be feasible. We introduce the \emph{slack variables} $\xi_i$ and $\xi^*_i$ for each point such that our problem becomes minimizing $\left(\frac{1}{2}|\bm{w}|^2+C\sum\limits_i(\xi_i+\xi_i^*)\right)$ with the constraints
\begin{eqnarray}
&&y_i-(\bm{w}\cdot\bm{x}_i+b)\leq\epsilon+\xi_i;\nonumber\\
&&(\bm{w}\cdot\bm{x}_i+b)-y_i\leq\epsilon+\xi^*_i;\nonumber\\
&&\xi_i,\xi_i^*\geq0.
\end{eqnarray}
The constant $C$ is the \emph{box constraint} which determines the trade-off between the flatness of $f$ and the amount up to which deviations larger than $\epsilon$ are tolerated \cite{svr}. This is depicted in Fig. \ref{softsvr}. 
Now we need to add
\begin{equation}
C\sum_i(\xi_i+\xi_i^*)-\sum_i(\eta_i\xi_i+\eta_i^*\xi_i^*)-\sum_i\alpha_i\xi_i-\sum_i\alpha_i^*\xi_i^*
\end{equation}
to (\ref{lagrangian}) and
\begin{eqnarray}
\frac{\partial L}{\partial\xi_i}&=&C-\eta_i-\alpha_i;\nonumber\\
\frac{\partial L}{\partial\xi_i^*}&=&C-\eta_i^*-\alpha_i^*
\end{eqnarray}
to (\ref{lagranderiv}).

Furthermore, one can easily imagine that we cannot always separate the points linearly\footnote{Examples can be found in Figure 6.1 in \cite{action}.}. This then requires the nonlinear SVMs. We will not discuss such cases here.

\subsection{Decision Trees}
\label{decisiontrees}

As the name suggests, this is a tree-like model for making decisions. Decision tree learning can be used to cope with both categorical and numerical data, and therefore comes the term CART (classicification and regression tree). There are also other advantages of decision trees. For instance, they are white box models so that the results can be explained by Boolean logic, unlike NNs.

\begin{figure}[h]
	\centering

	\tikzset{every picture/.style={line width=0.75pt}} 
	
	\begin{tikzpicture}[x=0.75pt,y=0.75pt,yscale=-1,xscale=1]
	
	\draw   (145.52,82.52) .. controls (145.52,76.16) and (150.68,71) .. (157.04,71) -- (241,71) .. controls (247.36,71) and (252.52,76.16) .. (252.52,82.52) -- (252.52,117.08) .. controls (252.52,123.44) and (247.36,128.6) .. (241,128.6) -- (157.04,128.6) .. controls (150.68,128.6) and (145.52,123.44) .. (145.52,117.08) -- cycle ;
	\draw    (275,48) -- (242.66,69.88) ;
	\draw [shift={(241,71)}, rotate = 325.91999999999996] [color={rgb, 255:red, 0; green, 0; blue, 0 }  ][line width=0.75]    (10.93,-3.29) .. controls (6.95,-1.4) and (3.31,-0.3) .. (0,0) .. controls (3.31,0.3) and (6.95,1.4) .. (10.93,3.29)   ;
	
	\draw    (345,48) -- (383.57,70.59) ;
	\draw [shift={(385.3,71.6)}, rotate = 210.35] [color={rgb, 255:red, 0; green, 0; blue, 0 }  ][line width=0.75]    (10.93,-3.29) .. controls (6.95,-1.4) and (3.31,-0.3) .. (0,0) .. controls (3.31,0.3) and (6.95,1.4) .. (10.93,3.29)   ;
	
	\draw   (275,8) -- (345,8) -- (345,48) -- (275,48) -- cycle ;
	\draw   (373.78,83.12) .. controls (373.78,76.76) and (378.94,71.6) .. (385.3,71.6) -- (500.26,71.6) .. controls (506.62,71.6) and (511.78,76.76) .. (511.78,83.12) -- (511.78,117.68) .. controls (511.78,124.04) and (506.62,129.2) .. (500.26,129.2) -- (385.3,129.2) .. controls (378.94,129.2) and (373.78,124.04) .. (373.78,117.68) -- cycle ;
	\draw    (385.3,129.2) -- (352.96,151.08) ;
	\draw [shift={(351.3,152.2)}, rotate = 325.91999999999996] [color={rgb, 255:red, 0; green, 0; blue, 0 }  ][line width=0.75]    (10.93,-3.29) .. controls (6.95,-1.4) and (3.31,-0.3) .. (0,0) .. controls (3.31,0.3) and (6.95,1.4) .. (10.93,3.29)   ;
	
	\draw    (500.26,129.2) -- (538.83,151.79) ;
	\draw [shift={(540.56,152.8)}, rotate = 210.35] [color={rgb, 255:red, 0; green, 0; blue, 0 }  ][line width=0.75]    (10.93,-3.29) .. controls (6.95,-1.4) and (3.31,-0.3) .. (0,0) .. controls (3.31,0.3) and (6.95,1.4) .. (10.93,3.29)   ;
	
	\draw   (343.3,160.2) .. controls (343.3,155.78) and (346.88,152.2) .. (351.3,152.2) -- (413.3,152.2) .. controls (413.3,152.2) and (413.3,152.2) .. (413.3,152.2) -- (413.3,184.2) .. controls (413.3,188.62) and (409.72,192.2) .. (405.3,192.2) -- (343.3,192.2) .. controls (343.3,192.2) and (343.3,192.2) .. (343.3,192.2) -- cycle ;
	\draw   (470.56,160.8) .. controls (470.56,156.38) and (474.14,152.8) .. (478.56,152.8) -- (540.56,152.8) .. controls (540.56,152.8) and (540.56,152.8) .. (540.56,152.8) -- (540.56,184.8) .. controls (540.56,189.22) and (536.98,192.8) .. (532.56,192.8) -- (470.56,192.8) .. controls (470.56,192.8) and (470.56,192.8) .. (470.56,192.8) -- cycle ;
	\draw    (157.04,128.6) -- (124.7,150.48) ;
	\draw [shift={(123.04,151.6)}, rotate = 325.91999999999996] [color={rgb, 255:red, 0; green, 0; blue, 0 }  ][line width=0.75]    (10.93,-3.29) .. controls (6.95,-1.4) and (3.31,-0.3) .. (0,0) .. controls (3.31,0.3) and (6.95,1.4) .. (10.93,3.29)   ;
	
	\draw    (241,128.6) -- (279.57,151.19) ;
	\draw [shift={(281.3,152.2)}, rotate = 210.35] [color={rgb, 255:red, 0; green, 0; blue, 0 }  ][line width=0.75]    (10.93,-3.29) .. controls (6.95,-1.4) and (3.31,-0.3) .. (0,0) .. controls (3.31,0.3) and (6.95,1.4) .. (10.93,3.29)   ;
	
	\draw   (53.04,159.6) .. controls (53.04,155.18) and (56.62,151.6) .. (61.04,151.6) -- (123.04,151.6) .. controls (123.04,151.6) and (123.04,151.6) .. (123.04,151.6) -- (123.04,183.6) .. controls (123.04,188.02) and (119.46,191.6) .. (115.04,191.6) -- (53.04,191.6) .. controls (53.04,191.6) and (53.04,191.6) .. (53.04,191.6) -- cycle ;
	\draw   (164.3,160.2) .. controls (164.3,155.78) and (167.88,152.2) .. (172.3,152.2) -- (281.3,152.2) .. controls (285.72,152.2) and (289.3,155.78) .. (289.3,160.2) -- (289.3,184.2) .. controls (289.3,188.62) and (285.72,192.2) .. (281.3,192.2) -- (172.3,192.2) .. controls (167.88,192.2) and (164.3,188.62) .. (164.3,184.2) -- cycle ;
	\draw    (172.3,192.2) -- (139.96,214.08) ;
	\draw [shift={(138.3,215.2)}, rotate = 325.91999999999996] [color={rgb, 255:red, 0; green, 0; blue, 0 }  ][line width=0.75]    (10.93,-3.29) .. controls (6.95,-1.4) and (3.31,-0.3) .. (0,0) .. controls (3.31,0.3) and (6.95,1.4) .. (10.93,3.29)   ;
	
	\draw    (281.3,192.2) -- (319.87,214.79) ;
	\draw [shift={(321.6,215.8)}, rotate = 210.35] [color={rgb, 255:red, 0; green, 0; blue, 0 }  ][line width=0.75]    (10.93,-3.29) .. controls (6.95,-1.4) and (3.31,-0.3) .. (0,0) .. controls (3.31,0.3) and (6.95,1.4) .. (10.93,3.29)   ;
	
	\draw   (68.3,223.2) .. controls (68.3,218.78) and (71.88,215.2) .. (76.3,215.2) -- (138.3,215.2) .. controls (138.3,215.2) and (138.3,215.2) .. (138.3,215.2) -- (138.3,247.2) .. controls (138.3,251.62) and (134.72,255.2) .. (130.3,255.2) -- (68.3,255.2) .. controls (68.3,255.2) and (68.3,255.2) .. (68.3,255.2) -- cycle ;
	\draw   (313.6,223.8) .. controls (313.6,219.38) and (317.18,215.8) .. (321.6,215.8) -- (383.6,215.8) .. controls (383.6,215.8) and (383.6,215.8) .. (383.6,215.8) -- (383.6,247.8) .. controls (383.6,252.22) and (380.02,255.8) .. (375.6,255.8) -- (313.6,255.8) .. controls (313.6,255.8) and (313.6,255.8) .. (313.6,255.8) -- cycle ;
	
	\draw (309,29) node  [align=left] {{\Large Jet}};
	\draw (168,46) node  [align=left] {Curved trajectory\\through Silicon Tracker};
	\draw (457,48) node  [align=left] {Straight trajectory\\through Silicon Tracker};
	\draw (199.02,99.8) node  [align=left] {Shower at EM \\calorimeter?};
	\draw (442.78,100.4) node  [align=left] {Shower at Hadron\\calorimeter?};
	\draw (353,136) node  [align=left] {Yes};
	\draw (534,133) node  [align=left] {No};
	\draw (378.3,172.2) node  [align=left] {Neurons};
	\draw (505.56,172.8) node  [align=left] {Photons};
	\draw (125,134) node  [align=left] {Yes};
	\draw (276,131) node  [align=left] {No};
	\draw (88.04,171.6) node  [align=left] {Electrons};
	\draw (226.8,172.2) node  [align=left] {Trajectory in\\muon chamber?};
	\draw (146,192) node  [align=left] {Yes};
	\draw (310,194) node  [align=left] {No};
	\draw (103.3,235.2) node  [align=left] {Muons};
	\draw (348.6,235.8) node  [align=left] {Pions};

	\end{tikzpicture}
	\caption{Particles in CMS.}
	\label{tree}
\end{figure}
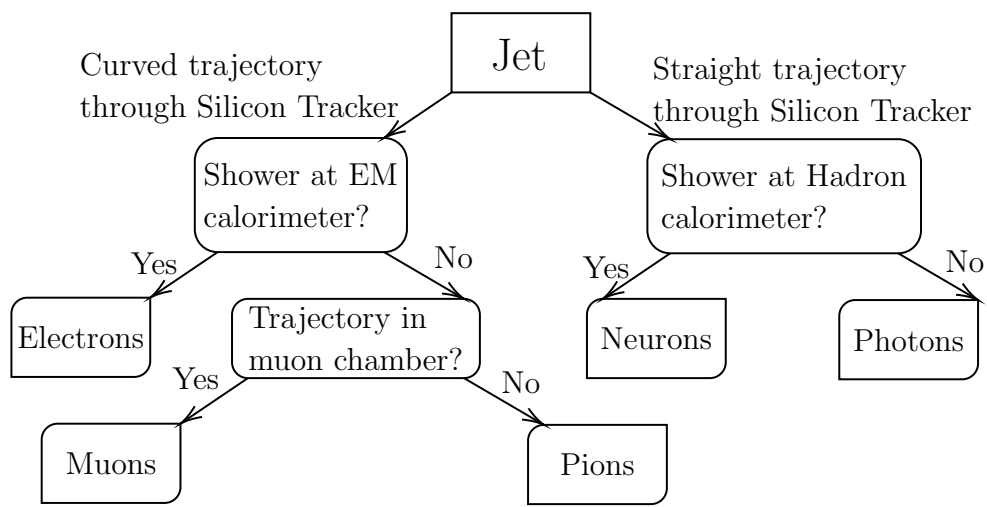

Suppose we are working at CERN and want to understand what particles are created in our experiment. We let the jet traverse the CMS so that different particles would stop in different detectors. The decision tree is then depicted in Fig. \ref{tree}.
This tree is drawn upside down with the \emph{root node} at the top. Internal nodes are known as \emph{decision nodes}, and the nodes at the end are called \emph{leaves} or \emph{terminal nodes}. We call the process of dividing a node (\emph{parent node}) into sub-nodes (\emph{child nodes}) a \emph{splitting}. A subsection of a tree is known as a \emph{branch}.

In general, we make a collection of rules based on our variables to get a best split of our data set. This yields the child nodes, and the same process is acted on each children node. As a result, this is a recursion procedure. Finally, the splitting will stop when no further gain can be made or we have some stopping rule preset.

The algorithms we use in decision trees are essentially greedy algorithms since a best choice is made at each decision. Different algorithms have different ways to measure the quality of the choice/splitting.

Often appeared in algorithms such as ID3 and C4.5, \emph{information gain} measures how ``best" our splitting is\footnote{Other algorithms may use Gini inpurity, variance reduction and so forth.}.
\begin{definition}
	An \emph{attribute} (aka \emph{feature}) is an individual measurable property or characteristic of a phenomenon being observed \cite{attribute}. For a dataset $D$ after split on an attribute $A$, the \emph{information gain} (aka \emph{mutual information}) $I(D,A)$ is
	\begin{equation}
	I(D,A)=S(D)-S(D|A),
	\end{equation}
	where $S(D)=-\sum\limits_{j\in J}p(j)\log_2p(j)$ is the Shannon entropy. We denote the set of classes as $J$, and $p(j)$ is the precentage of the number of elements in class $j$ in the number of total elements in $D$ such that $\sum\limits_{j\in J}p(j)=1$. Moreover, $S(D|A)$ is the conditional entropy such that
	\begin{equation}
	S(D|A)=-\sum_ap(a)\sum_jp(j|a)\log_2p(j|a)
	\end{equation}
	where $p(a)$ is the proportion of $A=a$ and $p(j|a)$ is the proportion $p(j)$ constrained by $A=a$ (cf. conditional probability).
\end{definition}
As entropy is used to quantify the uncertainty (or equivalently, our knowledge) of our information, the information gain, as the name suggests, measures the difference of entropy between before and after the dataset $D$ is split on attribute $A$. It shows how much uncertainty is reduced (or equivalently, how much knowledge we gain) after splitting. Now that we are making best choices, we often want the information gain to be maximized at each step.

In our particle jet example in Fig.\ref{tree}, there are four attributes: charge (charged or uncharged, determined by Sicicon tracker), (primarily) electromagnetic interaction (yes or no, determined by EM calorimeter), (primarily) nuclear interaction (yes or no, determined by hadron calorimeter) and absorption by calorimeters (yes or no, determined by muon chamber). For instance, let us compute the information gain of the first splitting based on charge. The Shannon entropy of the five types/classes of particles is\footnote{Here we only care about different kinds of particles, rather than the percentage of particle numbers of each kind in the jet.}
\begin{equation}
S(D)=-5\times\left(\frac{1}{5}\log_2\frac{1}{5}\right)=\log_25.
\end{equation}
There are three types of particles carrying charge, and two being neutral. Thus, the conditional entropy is
\begin{equation}
S(D|\text{Charge})=-\frac{3}{5}\times3\times\left(\frac{1}{3}\log_2\frac{1}{3}\right)-\frac{2}{5}\times2\times\left(\frac{1}{2}\log_2\frac{1}{2}\right)=\frac{3}{5}\log_23+\frac{2}{5}.
\end{equation}
Hence, the information gain is
\begin{equation}
I=\log_25-\left(\frac{3}{5}\log_23+\frac{2}{5}\right)=0.97.
\end{equation}

Decision trees are frequently used in machine learning and data mining. A main problem of decision trees is \emph{overfitting}. The algorithm may build a tree too close to the data, which leads to an overfitting tree. This is often over-complex and may give poor performance during validation and prediction. A technique known as \emph{pruning} is then used to reduce the size of learning trees. For more details, one is referred to \cite{action, datamining}.

\subsection{Types of Machine Learning}
\label{types}

So far we have been mainly talking about various models used in machine learning, there are different machine learning algorithms based on different types. Here, we will quickly introduce the three basic machine learning paradigms:

\begin{itemize}
	\item Supervised Learning: In supervised learning, our data is split into training and validation data. These data contain both inputs and outputs. Then we will train the machine with training data and examine its learning result using validation data. We are mainly using this approach in the next few sections, and a more detailed instruction is given in \ref{warmupWP4} by an example. The models of algorithms aforementioned (NNs, SVMs, decision trees etc.) are all commmonly used in supervised learning. It is worth mentioning that the \emph{no free lunch theorem}\footnote{This may reminds some readers of the famous saying by Alan Guth: `` The universe is the ultimate free lunch." Apparently, this refers to different contexts and topics from here. Although not everyone is a fan of inflation, we all agree that vacuum is full of fluctuations. It is kind of interesting to see a proverb quoted in various areas with different meanings.} tells us no learning algorithm would always work best on all supervised learning problems.
	
	\item Unsupervised Learning: In unsupervised learning, the algorithm is asked to find unknown patterns of the data without labels. Hence, the data only contains inputs. One usual approach in unsupervised learning is cluster analysis where the machine groups the data-points with similar properties. There is a category in machine learning which hybridize supervised and unsupervised learnings known as semi-supervised learning as well. Further discussions can be found in Part 3 of \cite{action}.
	
	\item Reinforcement Learning (RL): A reinforcement learning is usually modelled as a \emph{Markov decision process} (MDP), where
	\begin{definition}
		A Markov decision process is a tuple $(S,A,P,R,\gamma)$ where $S(\ni s)$ is a finite set of states and $A(\ni a)$ is a set of actions. Then $P$ is the \emph{state transition probability matrix} such that $P_{ss'}^a=\text{Prob}(S_{t+1}=s'|S_t=s,A_t=a)$ with $t$ labelling the time-step, and $R$ is the \emph{reward function} such that $R_s^a=E(R_{t+1}|S_t=s,A_t=a)$ with $E$ being the expectation value and $R_t$ being the reward at $t$. Moreover, $\gamma\in[0,1]$ is the \emph{discount factor}.
	\end{definition}
	As the MDP is built on the Markov chain, it inherits the memorylessness property. To ``reinforce" our learning, we not only give rewards to the \emph{agent} (which is the component making decisions of what action to take), but also accumulate the rewards:
	\begin{definition}
		A \emph{return} $G_t$ is the total discounted reward from $t$, viz, $G_t=\sum\limits_{k=0}^\infty\gamma^kR_{t+k+1}$.
	\end{definition}
	We see that discount factor is added since a reward received at present is more worthy than delayed rewards\footnote{In fact, creatures also seem to prefer immediate rewards.}. Therefore, the agent is able to know the value of being in a state $s$ in long term:
	\begin{definition}
		The \emph{state-value function} is $v_\pi(s)=E(G_t|S_t=s)$, and the \emph{action-value funtion} is $q_\pi(s,a)=E(G_t|S_t=s,A_t=a)$, where $\pi$ is the \emph{policy} such that $\pi(a|s)=\text{Prob}(A_t=a|S_t=s)$, which defines the behaviour of our agent.
	\end{definition}
	
	RL is now used not only in machine learning, but also in fields such as game theory, information theory, statistics and so forth. Such reinforcement learning with rewards reminds us of the Skinner box---we wish to explore this point in future works. Recently, RL is also applied to studying the string vacua as in \cite{RL}.
\end{itemize}

With different algorithms and categories in machine learning, what we need is an appropriate way to measure how well our machine learns. This is discussed in \S\ref{warmupWP4}.

\addcontentsline{toc}{section}{References}
\bibliographystyle{utphys}
\bibliography{references}
\end{document}